\documentclass[aps,pre,reprint,graphicx,notitlepage,amssymb,superscriptaddress,nobalancelastpage
,longbibliography
]{revtex4-1}
\usepackage[utf8]{inputenc}
\usepackage[T1]{fontenc}
\usepackage{lmodern}					
\usepackage{amsmath}
\usepackage{amsfonts}
\usepackage{amssymb,textcomp}
\usepackage[xcdraw]{xcolor}
\usepackage{colortbl}
\usepackage{tikz}
\usepackage{bm}
\usepackage{framed,xspace}
\usepackage{graphicx}
\usepackage{array,multirow}
\usepackage{natbib}
\usepackage[margin=2cm]{geometry}

\usepackage{hyperref}

\newcommand\beq{\begin{equation}}      
\newcommand\beqnn{\begin{eqnarray*}}   
\newcommand\beqa{\begin{eqnarray}}     
\newcommand\beqann{\begin{eqnarray*}}  

\newcommand\eeq{\end{equation}}        
\newcommand\eeqnn{\end{eqnarray*}}     
\newcommand\eeqa{\end{eqnarray}}       
\newcommand\eeqann{\end{eqnarray*}}    

\newcommand{\ket}[1]{\left| #1 \right\rangle}                                               
\newcommand{\bra}[1]{\left\langle #1\right|}                                               

\newcommand{\eq}[1]{\begin{align}#1\end{align}}
                                               
\def\nl {\nonumber \\}

\newcommand\bi{\begin{itemize}}
\newcommand\ei{\end{itemize}}

\def\nl {\nonumber \\}

\def\bv{\bm v}

\newcommand{\eref}[1]{(\ref{#1})}
\newcommand{\fref}[1]{Fig.~\ref{#1}}

\newcommand{\Eref}[1]{Eq.~(\ref{#1})}
\newcommand{\Sref}[1]{Sec.~\ref{#1}} 
\newcommand{\Fref}[1]{Figure~\ref{#1}}
\newcommand{\al}[1]{\begin{align} #1 \end{align}}

\def\z{\zeta}

\providecommand{\st}[1]{_{\text{#1}}}

\providecommand{\ut}[1]{^{\text{#1}}}

\def\onehalf{\frac{1}{2}}

\def\bra{\ensuremath{\langle}}
\def\ket{\ensuremath{\rangle}}
\def\eq{\st{eq}}

\def\const{\mathrm{const}}

\def\pd{\partial}

\def\b0{\bv{0}}

\def\ra{\rightarrow}

\def\Fcal{\mathcal{F}}

\def\Kcal{\mathcal{K}}

\def\Mcal{\mathcal{M}}

\def\Ocal{\mathcal{O}}
\def\Tcal{\mathcal{T}}

\def\res{\st{res}}

\def\gc{\ut{(gc)}}
\def\can{\ut{(c)}}
\def\cgc{\ut{(c,gc)}}
\def\br0{{(0)}}

\def\sgn{\mathrm{sgn}}
\def\tscal{x}
\def\amplPhit{\phi_t^\br0}
\def\amplPhimu{\phi_\mu^\br0}
\def\amplXipm{\xi_\pm^\br0}
\def\amplXip{\xi_+^\br0}
\def\amplXim{\xi_-^\br0}
\def\amplXimu{\xi_{\mu}^\br0}
\def\lenPhi0{l_\mden^\br0}
\def\lenH1{l_{h_1}^\br0}

\def\mass{\Phi}
\def\Mass{\Mcal}
\def\mden{\varphi}
\def\bcs{boundary conditions\xspace}
\def\tf{t^{\textrm f}}

\newcommand*\colvec[3][]{
    \begin{pmatrix}\ifx\relax#1\relax\else#1\\\fi#2\\#3\end{pmatrix}
}

\def\xcf{x_c^{\textrm f}}

\def\be{\begin{equation}}
\def\ee{\end{equation}}
\def\la{\langle}
\def\ra{\rangle}

\begin{document}

\title{Ensemble dependence of Critical Casimir Forces in Films with Dirichlet Boundary Conditions}
\author{Christian M. Rohwer}
\email[]{crohwer@is.mpg.de}
\affiliation{Max Planck Institute for Intelligent Systems, Heisenbergstr.\ 3, 70569 Stuttgart, Germany}
\affiliation{4th Institute for Theoretical Physics, Universit\"at Stuttgart, Pfaffenwaldring 57, 70569 Stuttgart, Germany}
\author{Alessio Squarcini}
\affiliation{Max Planck Institute for Intelligent Systems, Heisenbergstr.\ 3, 70569 Stuttgart, Germany}
\affiliation{4th Institute for Theoretical Physics, Universit\"at Stuttgart, Pfaffenwaldring 57, 70569 Stuttgart, Germany}
\author{Oleg Vasilyev}
\affiliation{Max Planck Institute for Intelligent Systems, Heisenbergstr.\ 3, 70569 Stuttgart, Germany}
\affiliation{4th Institute for Theoretical Physics, Universit\"at Stuttgart, Pfaffenwaldring 57, 70569 Stuttgart, Germany}
\author{S. Dietrich}
\affiliation{Max Planck Institute for Intelligent Systems, Heisenbergstr.\ 3, 70569 Stuttgart, Germany}
\affiliation{4th Institute for Theoretical Physics, Universit\"at Stuttgart, Pfaffenwaldring 57, 70569 Stuttgart, Germany}

\author{Markus Gross}
\email[]{gross@is.mpg.de}
\affiliation{Max Planck Institute for Intelligent Systems, Heisenbergstr.\ 3, 70569 Stuttgart, Germany}
\affiliation{4th Institute for Theoretical Physics, Universit\"at Stuttgart, Pfaffenwaldring 57, 70569 Stuttgart, Germany}

\date{\today}

\begin{abstract}
In a recent study [Phys. Rev. E \textbf{94}, 022103 (2016)] it has been shown that, for a fluid film subject to critical adsorption, the resulting critical Casimir force (CCF) may significantly depend on the thermodynamic ensemble.
Here, we extend that study by considering fluid films within the so-called ordinary surface universality class. We focus on mean-field theory, within which the order parameter (OP) profile satisfies Dirichlet boundary conditions and produces a nontrivial CCF in the presence of external bulk fields or, respectively, a nonzero total order parameter within the film.
Additionally, we study the influence of fluctuations by means of Monte Carlo simulations of the three-dimensional Ising model.
We show that, in the canonical ensemble, i.e., when fixing the so-called total mass within the film, the CCF is repulsive for large absolute values of the total OP, instead of attractive as in the grand canonical ensemble. Based on the Landau-Ginzburg free energy, we furthermore obtain analytic expressions for the order parameter profiles and analyze the relation between the total mass in the film and the external bulk field.
\end{abstract}

\maketitle

\section{Introduction}
\label{sec_intro}

Confining a critical fluid by parallel walls gives rise to a critical Casimir force (CCF) acting on the bounding surfaces \cite{fisher_wall_1978, krech_casimir_1994}. 
Here we consider fluids belonging to the Ising bulk universality class (UC), which, accordingly, are described by a one-component order parameter (OP) field $\phi$. 
The bulk UC splits up into several surface UCs, describing further universal properties induced by the surfaces \cite{diehl_field-theoretical_1986, diehl_theory_1997, brankov_theory_2000}. 
In a classical fluid, the constituent molecules are generically attracted towards an immersed solid surface. This attraction can be either strong or weak compared with the liquid-liquid interaction. Accordingly, for a one-component fluid the surfaces have a preference either for its liquid phase (in the case of a strong substrate) or the vapor phase (in the case of a weak substrate), whereas for a binary liquid mixture the walls attract that phase which is rich in the species preferred by the surfaces.
Near the critical point, this attraction gives rise to the phenomenon of critical adsorption, which, in the limit of infinitely strong adsorption (surface field $h_{1}\rightarrow\infty$), is described by the so-called \emph{normal} surface UC \cite{fisher_scaling_1981, liu_universal_1989, floter_universal_1995}. 
Fluids show also an enhanced molecular order near a solid surface \cite{diehl_field-theoretical_1986, floter_universal_1995}, which is modeled field-theoretically by a so-called surface enhancement parameter $c$.
The limit $c\to \infty$ (for finite adsorption strength $h_{1}$) defines the so-called \emph{ordinary} surface UC, in which the OP effectively satisfies Dirichlet \bcs.

While critical fluids are typically strongly adsorbed at container walls \cite{gambassi_critical_2009}, by suitable preparation of the surfaces it is nevertheless possible to approach the limit of weak adsorption, corresponding to the ordinary surface UC.
In Ref.\ \cite{nellen_tunability_2009}, this has been achieved by chemical treatment of the surface, while in Refs.\ \cite{sprenger_forces_2006, trondle_normal_2009, trondle_critical_2010, gambassi_critical_2011} surface patterning has been used.

The CCF stems from residual finite-size contributions of the free energy of the film. 
Remarkably, as has been shown in Refs.\ \cite{gross_critical_2016, gross_statistical_2017}, the amplitude and the scaling function of the CCF depend not only on the bulk and the surface UC, but also on the thermodynamic ensemble under consideration.
In fact, CCFs are typically studied for fluid films which can exchange particles with their environment---a situation which realizes the grand canonical ensemble.
However, global OP conservation, which is applicable for the canonical ensemble, can induce drastic changes of the CCF \cite{gross_critical_2016, gross_statistical_2017}.
Hitherto, only a few studies have focused on the effect of a global OP constraint on the critical behavior \cite{eisenriegler_helmholtz_1987, brankov_probabilistic_1989, blote_three-dimensional_2000, caracciolo_finite-size_2001, pleimling_crossing_2001}.
In the present study, building on Ref.\ \cite{gross_critical_2016} (where critical adsorption has been investigated), we consider Ising-type fluid films within the ordinary surface UC, subject to a global OP constraint. 
We focus on mean-field theory, within which the effects of fluctuations are neglected and the CCF is a consequence of the presence of a spatially varying OP profile across the film.

In the grand canonical ensemble, a nonzero external bulk field $\mu$ acting in the film does induce a nontrivial OP profile.
In the canonical ensemble, instead, a nonzero value $\mass$ of the total integrated OP, henceforth called the \emph{mass}, is imposed:
\beq  
\mass \equiv  A \int_{0}^{L} d z\,\phi(z).
\label{eq_Mass0}
\eeq 
Here $A$ denotes the transverse area of the film, $L$ its thickness, and $z$ the associated transverse coordinate.
We generally assume the film to be homogeneous in the remaining, lateral directions. Henceforth we consider all extensive quantities, such as $\mass$, as quantities \emph{per transverse area} $A$, i.e., $\int_{0}^{L}d z \, \phi(z)$.
We find that the OP constraint in \Eref{eq_Mass0} can change, \emph{inter alia}, the character of the CCF from attractive in the grand canonical case to repulsive in the canonical case.

In passing, we recall that, for a critical fluid film within the ordinary surface UC, the critical temperature $T_c$ is shifted from its bulk value $T_c\ut{b}$ to $T_c\ut{f}<T_c\ut{b}$.
For Dirichlet \bcs and vanishing external fields $\mu=0$, the OP profile vanishes above the film critical point, i.e., for temperatures $T> T_c\ut{f}$.
CCFs for Ising-type systems in the ordinary surface UC (including crossover effects to the normal surface UC) have been previously studied within the grand canonical ensemble in Refs.\ \cite{krech_finite-size_1991, krech_free_1992, krech_specific_1992, schmidt_crossover_2008, hasenbusch_thermodynamic_2011, diehl_critical_2011, mohry_crossover_2010, vasilyev_critical_2011, vasilyev_critical_2013}.

In \Sref{sec_prelims}, we define the general scaling variables required for the description of the universal critical properties and outline the scaling relations expected for the OP profile. We furthermore introduce the Landau-Ginzburg model which is analyzed in the remaining part of this study.
The OP profile resulting from the Landau-Ginzburg model within mean-field theory is determined perturbatively in \Sref{MFTeq} and fully via numerical studies in \Sref{sec_MFT_comp}. The associated relation between the total mass and the external bulk field is analyzed separately in \Sref{sec_phasediag}.
In \Sref{sec_Forces}, the CCF is studied analytically within linearized MFT and numerically within full MFT, focusing on ensemble differences. In \Sref{sec_MC} the predictions of MFT are compared to Monte Carlo (MC) simulations of the three-dimensional Ising model.

\section{Preliminaries}
\label{sec_prelims}

\subsection{Scaling behavior}
\label{sec_scaling}

Here, we summarize the general scaling behavior expected for the OP profile and the CCF in a $d$-dimensional film of thickness $L$.
In the following we focus on the so-called ordinary fixed point, at which $c=\infty$ and, accordingly, the dependence of the scaling functions on $c$ drops out.
The following finite-size scaling relations apply to isotropic systems with short-ranged interactions below the upper critical dimension $d=4$ of the Ising universality class \cite{privman_universal_1984,privman_finite-size_1990}.
The universal properties of a critical film are  expected to be controlled by the following set of scaling variables:
\begin{subequations}\begin{alignat}{2} 
	\zeta &\equiv z/L, \label{eq_zeta}\\
	\tscal &\equiv  \left(\frac{L}{\amplXip}\right)^{1/\nu}t,\label{eq_tscal}\\
	B  &\equiv  \left(\frac{L}{\amplXimu}\right)^{\Delta/\nu} \mu, \label{eq_Bscal}\\
	\Mass &\equiv \left(\frac{L}{\amplXip}\right)^{\beta/\nu} \frac{\mden}{\amplPhit} ,
	\label{eq_Mass}
	\end{alignat}\label{eq_scalvar}\end{subequations} 
where
\beq \mden \equiv \frac{\Phi}{L}
\label{eq_mden}\eeq 
is the mean mass density of the film, $\beta$, $\nu$, and $\Delta$ are standard bulk critical exponents, and
\beq t = \frac{T-T_c\ut{b}}{T_c\ut{b}}
\label{eq_tred}\eeq
is the reduced temperature relative to the bulk critical temperature $T_c\ut{b}$.
In the case of a one-component fluid the external bulk field $\mu$ describes the deviation of the chemical potential from its critical value in the bulk, while for a binary liquid mixture, $\mu$ represents the deviation of the difference in the chemical potentials of the two species A and B from its bulk critical value: $\mu\equiv (\mu_A-\mu_B) - (\mu_{A,c}-\mu_{B,c})$.
The quantities $\amplXip$ and $\amplXimu$ (as well as $\amplXim$, which we include here for completeness) denote non-universal amplitudes defined in terms of the (bulk) correlation length $\xi_t$ at zero bulk field and $\xi_\mu$ at zero reduced temperature:
\begin{subequations}\begin{alignat}{3} 
 \xi_t &= \amplXipm |t|^{-\nu},\qquad &&\text{for }\mu=0\text{ and }t\to 0^\pm,\label{eq_gen_correlt}\\
 \xi_\mu &= \amplXimu |\mu|^{-\nu/\Delta},\qquad &&\text{for }t=0\text{ and }\mu\to 0. \label{eq_gen_correlmu}
\end{alignat}\label{eq_gen_correl}\end{subequations}
The value of $\amplXipm$ is different for $t\lessgtr 0$, but the amplitude ratio $U_\xi\equiv \amplXip/\amplXim$ forms the universal number $U_\xi \simeq 1.9$ in $d=3$ and $U_\xi = \sqrt{2}$ in $d=4$ spatial dimensions \cite{pelissetto_critical_2002}.
Except for Sec.~\ref{sec_MFT_comp}, we focus on the supercritical regime and therefore in the scaling relations we use solely $\amplXip$.
The non-universal amplitude $\amplPhit$ is defined in terms of the bulk OP $\phi_b$, which, near criticality, behaves as
\begin{subequations}\begin{alignat}{3}
 \phi_{b,t} &= \theta(-t)\amplPhit |t|^\beta, \qquad &&\text{for }\mu=0 \text{ and }t\to 0,\label{eq_gen_phit}\\
 \phi_{b,\mu} &= \sgn(\mu) \amplPhimu |\mu|^{1/\delta}, \qquad &&\text{for }t=0\text{ and }\mu\to 0, \label{eq_gen_phimu}
\end{alignat}\label{eq_gen_phi}\end{subequations}

\begin{widetext}
in the case of a vanishing external field $\mu$ and a vanishing reduced temperature $t$, respectively.

The OP profiles in the grand canonical and the canonical ensemble fulfill the following scaling relations \cite{binder_critical_1983, diehl_field-theoretical_1986, privman_finite-size_1990, brankov_theory_2000, krech_casimir_1994}:
\begin{subequations}
\begin{align}
 \phi\gc(z, t, \mu, L) 
    &= \amplPhit \left(\frac{L}{\amplXip}\right)^{-\beta/\nu} m\gc\left(\frac{z}{L}, \left(\frac{L}{\amplXip}\right)^{1/\nu}t, \left(\frac{L}{\amplXimu}\right)^{\Delta/\nu} \mu  \right), \label{eq_gen_prof_scal}\\
 \phi\can(z, t, \mden, L) 
    &= \amplPhit \left(\frac{L}{\amplXip}\right)^{-\beta/\nu}  m\can\left( \frac{z}{L}, \left(\frac{L}{\amplXip}\right)^{1/\nu}t, \left(\frac{L}{\amplXip}\right)^{\beta/\nu} \frac{\mden}{\amplPhit} \right), \label{eq_gen_prof_scal_c}
\end{align}
\end{subequations}
\end{widetext}
where $m\cgc$ are the corresponding universal scaling functions.
In order to simplify the notation, we henceforth drop the superscripts (c) and (gc) on $\phi$ and $m$.
The scaling variable $\Mass$ in Eq.\ \eqref{eq_Mass} is related to the scaling function $m$ via
\beq \Mass =\int_{0}^{1} d\zeta\, m(\zeta).
\label{eq_Mass_dimless}\eeq 
The general scaling behavior of the CCF is discussed in Sec.~\ref{sec_Forces}.
We remark that the scaling relations stated above apply for simple fluids with isotropic short-ranged interactions, so that two-scale factor universality holds. For a discussion of the influence of anisotropy as well as of long-ranged (van der Waals) interactions on the critical behavior we refer to Refs.\ \cite{dantchev_universality_2003,chen_nonuniversal_2004,dantchev_interplay_2007,dohm_diversity_2008,diehl_dynamic_2009,selke_critical_2009,dohm_crossover_2018}.

\subsection{Model and boundary conditions}
\label{sec_model}

We aim at determining the order parameter profile between two parallel plates, located at $z=0,L$ and subject to the constraint of a specified total mass $\mass$ [see \Eref{eq_Mass0} and recall that here and in the following $\mass$ is considered per area $A$]. 
The canonical Landau-Ginzburg (LG) free energy functional for \textit{f}ilms, in units of $k_B T$ per transverse area $A$ of the plates, is given by
\al{
\Fcal_f\can[\phi] \equiv &\int_{0}^{L} dz \left[\onehalf (\pd_z \phi)^2 + \onehalf \tau \phi^2 +\frac{1}{4!} g \phi^4 \right] \nl
&+  \left[ c_1\phi^2(z=0)+ c_2 \phi^2(z=L) \right].
\label{eq_Landau_func_c}
}
The integral represents the bulk contribution, whereas the terms $\propto c_1,c_2$ are surface enhancements giving rise to Robin-type boundary conditions ~\cite{diehl_field-theoretical_1986} on $\phi$ --- see \Eref{eq_ELE0_bc} below.
Within MFT, the coupling constants $\tau$ and $g$ are given by $\tau=(\amplXip)^{-2} t$ and $g=6(\amplXip \amplPhit)^{-2}$, where $t$ is the reduced temperature [\Eref{eq_tred}] and the amplitudes $\amplXip$ and $\amplPhit$ are defined in Eqs.~\eqref{eq_gen_correlt} and \eqref{eq_gen_phit}. Within MFT, one has $\amplXim/\amplXip = 1/\sqrt{2}$.
Equilibrium states minimize \Eref{eq_Landau_func_c}, subject to the constraint in \Eref{eq_Mass0}. 
In the grand canonical ensemble the LG functional for \textit{f}ilms (per $k_B T$ and area $A$) reads
\begin{widetext}
\begin{equation} \Fcal_f\gc([\phi];\mu) \equiv \int_{0}^{L} dz \left[\onehalf (\pd_z \phi)^2 + \onehalf \tau \phi^2 +\frac{1}{4!} g \phi^4 - \mu\phi\right] +  \left[ c_1\phi^2(z=0)+ c_2\phi^2(z=L) \right],
\label{eq_Landau_func_gc}
\end{equation}
which is to be minimized with respect to $\phi$, taking for the external bulk field (i.e., the chemical potential) $\mu$ a value such that Eq.~\eqref{eq_Mass0} is obeyed. Minimization of the grand canonical energy functional leads to the Euler-Lagrange equation (ELE)
\beq 
\pd_z^2 \phi - \tau \phi - \frac{g}{6}\phi^3 + \mu = 0,
\label{eq_ELE0}
\eeq
subject to the boundary conditions
\beq \begin{split}
	\pd_z \phi\big|_{z=0} = c_1 \phi(z=0), \qquad
	\pd_z\phi\big|_{z=L} = -c_2 \phi(z=L),
\end{split}
\label{eq_ELE0_bc}
\eeq 
induced by the surface enhancement terms. In what follows, we shall study the limits $c_1,c_2\to\infty$, for which Dirichlet boundary conditions $\phi(z=0)=0=\phi(z=L)$ emerge.

Within MFT, the finite-size scaling variables defined in \Eref{eq_scalvar} turn into
\beq 
\tscal = L^{2} \tau,\quad 
B = \sqrt{\frac{g}{6}} L^{3} \mu,\quad
m(\zeta) = \sqrt{\frac{g}{6}} L \phi(\zeta L),\quad
\varphi = \Phi/L,\quad\text{and}\quad
\Mass = \sqrt{\frac{g}{6}} L \mden,
\label{eq_scalvar_mft}
\eeq  
in terms of which $\Fcal_f\gc$ in Eq.~\eqref{eq_Landau_func_gc} can be expressed as
\beq 
\Fcal_f\gc([m];B) =\frac{\Delta_0}{L^3} \left\{\int_{0}^{1} d\zeta \left[\onehalf ( m')^2 + \onehalf \tscal  m^2 +\frac{1}{4}  m^4 - B  m\right] 
+ \left[c_1 m^2(0)+c_2 m^2(1)\right] \right\}.
\label{eq_Landau_func_ndim}
\eeq 
\end{widetext}
The non-universal amplitude $\Delta_{0}$ is given by
\beq \Delta_0 \equiv \left(\amplXip \amplPhit\right)^2 = \frac{6}{g}
\label{eq_Delta0}\eeq 
in terms of the amplitudes of the correlation length and the bulk OP [see Eqs.\ \eqref{eq_gen_phit} and \eqref{eq_gen_correlt}].
We note that $\Delta_0$ has the same dimension as $L^{4-d}$, while the film free energies in Eqs.\ \eqref{eq_Landau_func_c} and \eqref{eq_Landau_func_gc}, being defined per area $A$, have the dimension of $1/L^{d-1}$.
The dimensionless form of the ELE, following from Eqs.~\eqref{eq_ELE0} and \eqref{eq_ELE0_bc}, reads
\beq  
m''(\zeta) - \tscal  m(\zeta) -  m^3(\zeta) + B = 0,
\label{eq_ELE}
\eeq
with the corresponding Dirichlet boundary conditions (obtained in the limits $c_1,c_2\to\infty$)
\beq 
m(0) = m(1) = 0.
\label{eq_ELE_BC}
\eeq 
Equations~\eref{eq_ELE} and \eref{eq_ELE_BC} are independent of the plate separation $L$ and the coupling constant $g$, because these variables can be scaled out such that they appear as prefactors in Eq.~\eqref{eq_Landau_func_ndim}. 
In general, the dimensionless counterpart of $g$ is fixed under renormalization-group flow, which requires to include fluctuations into the theory. 
Within MFT, $g$ and $\Delta_0$ can be related to experimentally accessible critical amplitudes via \Eref{eq_Delta0}.

\section{Perturbative Mean Field Analysis}
\label{MFTeq}

In order to make analytical progress, we  address the nonlinear term of the ELE in Eq.~\eref{eq_ELE} perturbatively by introducing a parameter $\epsilon$ (eventually to be set to unity):
\al{
m''(\z) - x m(\z)  - \epsilon m^3(\z) + B = 0.
\label{ELE_MFT}
}
This equation must be solved subject to the Dirichlet boundary conditions in \Eref{eq_ELE_BC} and under the constraint [\Eref{eq_Mass_dimless}]
\al{
\int_{0}^{1} d\z\,m(\z) = \mathcal M.
\label{MassConstr}
}
In a first step, we solve Eq. \eref{ELE_MFT} without this constraint by carrying out perturbation theory in terms of powers of $\epsilon$, with the series expansions
\al{
m &= \sum_{i \geqslant 0} \epsilon^{i} m_{i} = m_{0} + \epsilon m_{1} + \epsilon^{2} m_{2} + \dots \, ,\nl
B &= \sum_{i \geqslant 0} \epsilon^{i} B_{i} = B_{0} + \epsilon B_{1} + \epsilon^{2} B_{2} + \dots \, .
\label{epsexpand}
}
The boundary conditions from Eq.~\eref{eq_ELE_BC} hold for each term $i$. Concerning the expansion of the mass constraint in Eq.~\eref{MassConstr}, we choose 
\al{ \mathcal M_0 = \mathcal{M},\quad \mathcal M_{i\geq 1} = 0,
\label{constexp}
}
where $\mathcal M_i =\int_{0}^{1} d\z\,m_i(\z)$.

\begin{figure}[t]
\centering
\includegraphics[width=.99\columnwidth]{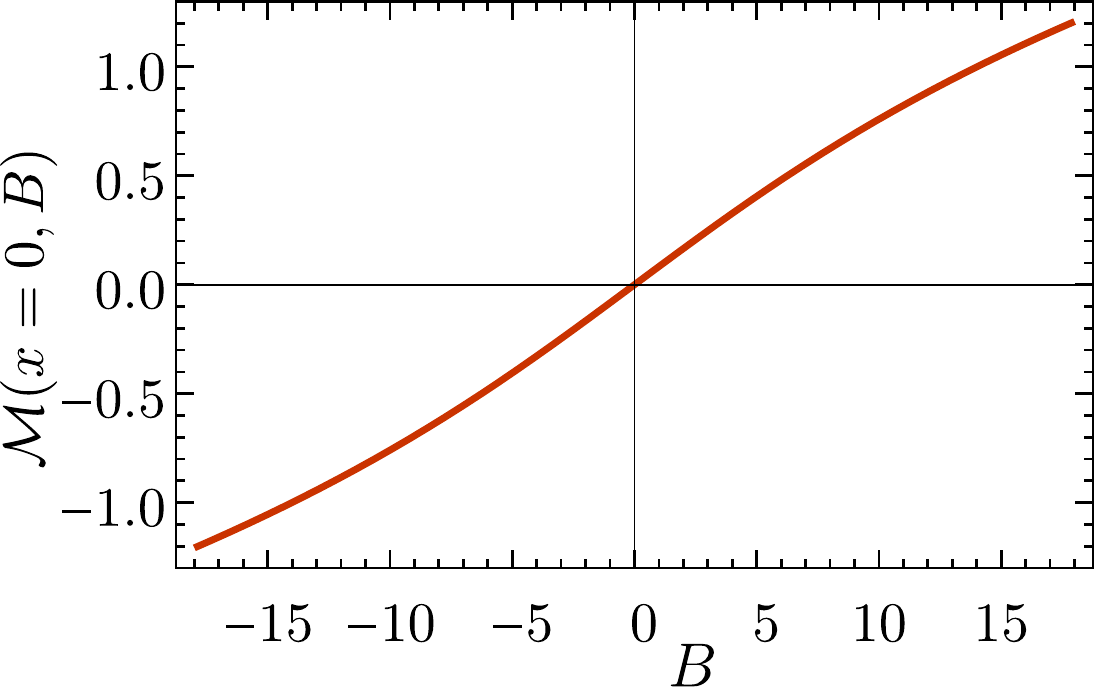}
\caption{The total mass as a function of the bulk field $B$ at the bulk critical temperature ($x=0$), determined from the (non-perturbative) numerical solution of the unconstrained ELE in Eq.~\eref{ELE_MFT}.}
\label{mass0}
\end{figure}

As a side remark, one infers from the structure of the ELE that, if $m(\zeta)$ is a solution of Eq. (\ref{ELE_MFT}) with parameters $x$ and $B$, then $-m(\zeta)$ will be a solution for the parameters $x$ and $-B$. Thus, the total mass $\mathcal{M}(x,B)$ is an odd function of the bulk field $B$, i.e., $\mathcal{M}(x,B) = - \mathcal{M}(x,-B)$. This feature is illustrated in Fig.~\ref{mass0} for the full, numerical (non-perturbative) solution of Eq.~\eref{ELE_MFT}, which must hold also at each perturbative order.

\subsection{Solution at $\mathcal O (\epsilon^0)$}
At this order, Eq.~\eref{ELE_MFT} yields
\al{
m''_0 = x m_0- B_0,
}
with the solution
\al{
m_0(\z) = \frac{B_0}{x} \left[1- \text{sech}\left(\frac{\sqrt{x}}{2}\right) \cosh \left((\zeta-1/2)  \sqrt{x}\right) \right].
\label{m0}
}
In contrast to the case of critical adsorption considered in Ref.~\cite{gross_critical_2016}, the lowest order MFT solution for Dirichlet boundary conditions is well-behaved near the bulk critical point. This is revealed by a series expansion for small $x$, yielding $m_0(\z) \simeq \frac{1}{8} [B_0-4 B_0 (\zeta-1/2) ^2] $. 
By using \Eref{eq_scalvar_mft}, \Eref{m0} can be written in terms of dimensional variables:
\al{
\phi_0(z) = \frac{\mu}{\tau} \big[1 - \text{sech}\left(\frac{L \sqrt{\tau }}{2}\right) \cosh \left( \sqrt{\tau } (z-\frac L 2)\right)\big],
\label{eq_phi0z}
}
which will be useful for the analysis presented in Sec.~\ref{sec_Forces_ST}.

Implementing now the constraint in Eq.~\eref{constexp} selects and fixes, at this order, the value $B_0= \tilde B_0$:
\al{
\tilde B_0 &= \frac{\mathcal M x^{3/2}}{\sqrt{x}-2 \tanh \left(\frac{\sqrt{x}}{2}\right)}\;\;\to\begin{cases}
 12\mathcal M ,\quad&x\to 0 \, ,\\
 \mathcal M x ,\quad&x\to\infty \, ,
\end{cases}
\label{B0const}
}
where the last expression exhibits the asymptotic scaling behavior close to the bulk critical point and for thick films, respectively. Inserting Eq.~\eref{B0const} into Eq.~\eref{m0} gives the contribution to the constrained order parameter at this order:
\al{
\tilde m_0(\z) = \frac{\mathcal M \sqrt{x} \left[1-\text{sech}\left(\frac{\sqrt{x}}{2}\right) \cosh \left((\zeta-1/2)  \sqrt{x}\right)\right]}{\sqrt{x}-2 \tanh \left(\frac{\sqrt{x}}{2}\right)}.
\label{m0tilde}
}
The asymptotic scaling of this expression,
\al{
\tilde m_0(\z) \to
\begin{cases}
 3\mathcal M(\frac 1 2 - 2 (\z-\frac{1}{2})^2),\quad&x\to 0,\\
 \mathcal M,\quad&x\to\infty,
\end{cases}
}
shows that at bulk criticality the lowest order MFT contribution for Dirichlet boundary conditions is a parabolic profile. In turn, away from criticality, the (spatially constant) solution must vanish due to the boundary conditions, which shows that $\mathcal M\to0$ if $x\to\infty$. Consequently, $\tilde B_0$ in Eq.  \eref{B0const} must also vanish away from criticality. Finally, expressing \Eref{m0tilde} in terms of dimensional variables, one finds the constrained profile
\al{
\tilde \phi_0(z)= \varphi \frac{1-\text{sech}\left(\frac{L \sqrt{\tau }}{2}\right) \cosh \left(\sqrt{\tau } (z-L/2)\right)}{1-\frac{2 \tanh \left(\frac{L \sqrt{\tau }}{2}\right)}{L \sqrt{\tau }}},
\label{eq_phi0ztilde}
}
which indeed satisfies the relation $\int_0^Ldz\;\tilde \phi_0(z) =\varphi L =\Phi$.

\subsection{Solution at $\mathcal O (\epsilon^1)$}
To linear order in $\epsilon$, Eq.~\eref{ELE_MFT} gives
\al{
m''_1 = x m_1+ m_0^3   - B_1.
}
The solution of this differential equation vanishes in the limit $B\to 0$. (The full expression is cumbersome and is not shown here.) Implementing the constraint of Eq.~\eref{constexp}, one finds the following corresponding specific expression $B_{1} = \tilde B_{1}$:
\al{
\tilde B_1 &= 
\frac{{\tilde B_0}^3 \text{sech}^4\left(\frac{\sqrt{x}}{2}\right)}{48 x^3 \left(\sqrt{x}-2 \tanh \left(\frac{\sqrt{x}}{2}\right)\right)} \nl
&\quad\times \Big[108 \sqrt{x}-160 \sinh \left(\sqrt{x}\right)-25 \sinh \left(2 \sqrt{x}\right)\nl
&\qquad +96 \sqrt{x} \cosh \left(\sqrt{x}\right)+6 \sqrt{x} \cosh \left(2 \sqrt{x}\right)\Big],
}
which exhibits the asymptotic scaling behavior
\al{
\tilde B_1 \to
\begin{cases}
 \frac{72}{35}\mathcal M^3 ,\quad&x\to 0,\\
 \mathcal M^3,\quad&x\to\infty.
\end{cases}
}
From this the constrained profile for very small and very large $x$ can be calculated:
\al{
\tilde m_1(\z) \to
\begin{cases}
 -\frac{9 \mathcal M^3}{8960} \scriptstyle{ \big[3840 (\zeta-1/2) ^8} \\ 
 \scriptstyle{ \;\;-5376 (\zeta-1/2) ^6  +3360 (\zeta-1/2) ^4} \\
 \scriptstyle{ \;\;-656 (\zeta-1/2) ^2+23\big]} ,&x\to 0,\\
 0,\quad&x\to\infty.
\end{cases}
}
At bulk criticality, a polynomial solution obeying the boundary conditions in \Eref{eq_ELE_BC} is obtained. As it was the case for the contribution $\mathcal O(\epsilon^0)$, the constrained profile vanishes away from criticality. 

The perturbative solution of the ELE to $\Ocal(\epsilon^2)$ is reported in Appendix \ref{app1}.

\section{Comparison of perturbative MFT solutions with exact and numerical results}
\label{sec_MFT_comp}

In this section we compare the leading perturbative solution at order $\mathcal{O}(\epsilon^{0})$ with numerical solutions of the full, nonlinear ELE (\ref{ELE_MFT}). In the case of zero external field, the full solution $m(\zeta)$ can be computed analytically (see Sec.~\ref{B0solution} below). In Sec.~\ref{unconstrprof} we consider the unconstrained solution $m(\zeta)$ for a given pair of parameters $(x,B)$, and compare it with $m_{0}(\zeta)$ given by Eq.~(\ref{m0}). Therefore, in \Sref{constrprof} we impose the constraint on the total mass and regard the corresponding solution $\tilde{m}(\zeta)$ as a function of the independent parameters $(x,\mathcal{M})$. The latter is compared with $\tilde{m}_{0}(\zeta)$ as given by Eq.~(\ref{m0tilde}).

\subsection{Exact analysis for $B=0$ and location of the film critical point}
\label{B0solution}
\begin{figure}[t]
\centering
\includegraphics[width=.99\columnwidth]{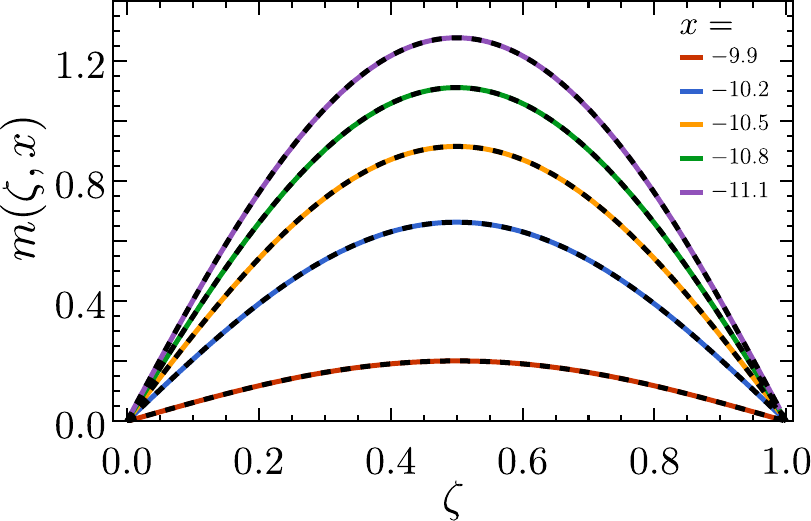}
\caption{OP profiles $m(\zeta,x)$ across the film in the grand canonical ensemble obtained for zero scaled bulk field, $B=0$. The exact result from Eq.~(\ref{m_exact}) (colored lines) is compared with the  numerical solutions (black dashed lines) of the nonlinear MFT for several values of $x$ below the film critical point, i.e., in the range $x< x_c\ut{f}$ [see \Eref{eq_x_c_film}].}
\label{exact-num}
\end{figure}

For $B=0$ an exact expression for the order parameter profile can be obtained in closed form in terms of elliptic functions \cite{gambassi_critical_2006}. According to Eq. (\ref{eq_ELE}), the associated ELE is
\begin{equation}
\label{ELE_B0}
m^{\prime\prime}(\zeta) - x m(\zeta) - m^{3}(\zeta) = 0 ,
\end{equation}
subject to the boundary conditions $m(0)=m(1)=0$. 
Beside the trivial solution $m(\zeta)=0$, there is a non-vanishing solution for $x\leqslant x_c\ut{f}$, where
\beq 
x_c\ut{f} = -\pi^{2}\, \simeq -9.87
\label{eq_x_c_film}\eeq 
denotes the scaled reduced temperature (relative to the bulk critical point) of the \emph{film} critical point. 
MC simulations of the Ising model \cite{vasilyev_universal_2009} yield a value $x_c\ut{f} \simeq -7.6$ for the film critical point, while field theoretic renormalization group studies \cite{dohm_pronounced_2014} predict $x_c\ut{f} \simeq -6.44$.
Here and in the following, when considering the regime $t<0$, i.e., $x<0$, we define $x$ as
\begin{equation}
\label{eq_x_neg}
x = \left( \frac{L}{\xi_{-}^{(0)}} \right)^{1/\nu} \frac{T-T_{\textrm{c}}^{\textrm{b}}}{T_{\textrm{c}}^{\textrm{b}}} \, .
\end{equation}
One finds that
\begin{equation}
\label{m_exact}
m_{\textrm{exact}}(\zeta) = 2\sqrt{2} k K(k^{2}) \, \textrm{sn}\left(2K(k^{2})\zeta;k^{2}\right) ,
\end{equation}
where
\begin{equation}
\label{elliptic_integral}
K(k^{2}) = \int_{0}^{1}\frac{d u}{\sqrt{(1-u^{2})(1-k^{2}u^{2})}} 
\end{equation}
is the complete elliptic integral of the first kind, $k$ is the elliptic modulus, determined implicitly by $x$ through $x = - 4 K^{2}(k^{2}) (1+k^{2})$, and $\textrm{sn}$ is Jacobi's \emph{elliptic sine} 
(see Refs.\ \cite{gradshteyn_table_2014, olver_nist_2010} for more details). 
 As shown in Fig.~\ref{exact-num}, the numerical solution of Eq. (\ref{ELE_B0}) perfectly matches the exact solution given in Eq. (\ref{m_exact}).

\subsection{Unconstrained profiles} 
\label{unconstrprof}

\subsubsection{Profiles for $x  = 0$}

\begin{figure}[t]
\centering
\includegraphics[width=.99\columnwidth]{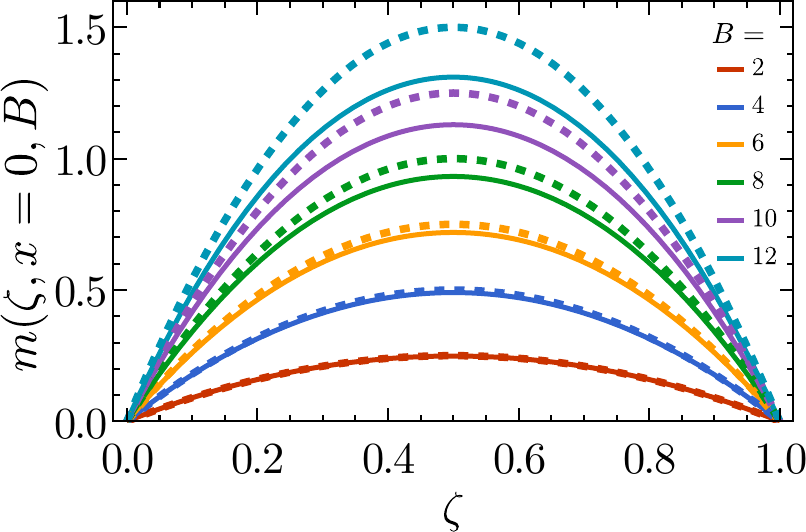}

\caption{The numerical solution $m(\zeta)$ (full curves) of the nonlinear MFT in the grand canonical ensemble is compared with $m_{0}(\zeta)$ (\Eref{m0}; dashed curves) for $x=0$ and $B\in\{2,4,6,8,10,12\}$. The discrepancy between the numerical and the truncated perturbative results grows as $B$ increases, and its maximum occurs at the midpoint $\zeta=1/2$ of the film.}
\label{uncontrained_profiles}
\end{figure}
In Fig.~\ref{uncontrained_profiles}, a comparison is shown of the unconstrained profiles obtained numerically with the perturbative approach at leading order. The perturbative solution $m_{0}(\zeta)$ [\Eref{m0}] deviates significantly from the numerical solution for large values of the bulk field $B$, with the largest deviations being localized in the middle of the film (i.e., $\zeta=1/2$). 
Close to the film boundaries at $\zeta=0,1$, the inaccuracy of the perturbative solution is mitigated by the fact that the Dirichlet boundary conditions are satisfied for all values of $B$.

\subsubsection{Profiles for $x \neq 0$} 
The approach outlined above can be followed also for $x\neq0$, and in principle the entire phase diagram can be explored. However, the same qualitative behavior encountered for $x=0$ occurs also for $x\neq0$. In general, the strongest inaccuracy is observed for $x<0$ (as the phase-separating regime is approached) and for large values of $B$ (where nonlinear effects become more dominant due to the term $\propto m^{3}$ in the ELE).

\subsection{Constrained profiles at $x=0$ and $x=x_{c}^{\mathrm{f}}$}
\label{constrprof}

\begin{figure}[t]
\centering
\includegraphics[width=.99\columnwidth]{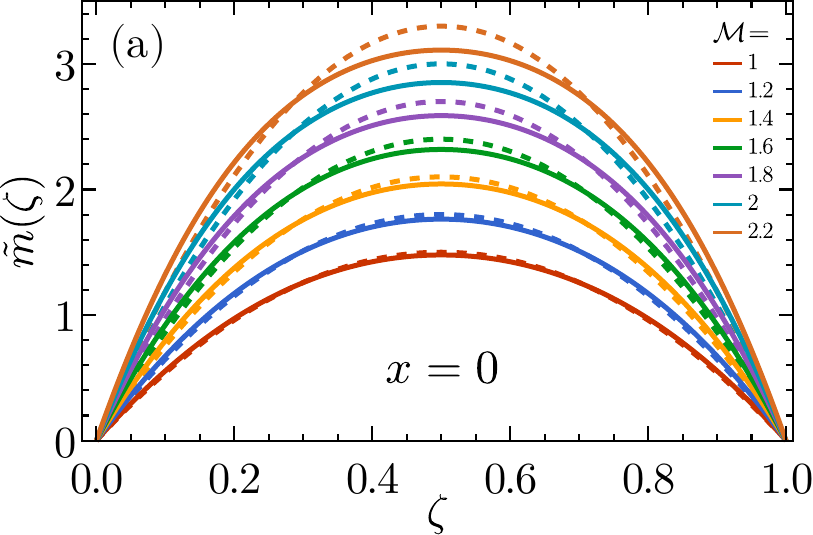}
\hspace{.2cm}

\includegraphics[width=.99\columnwidth]{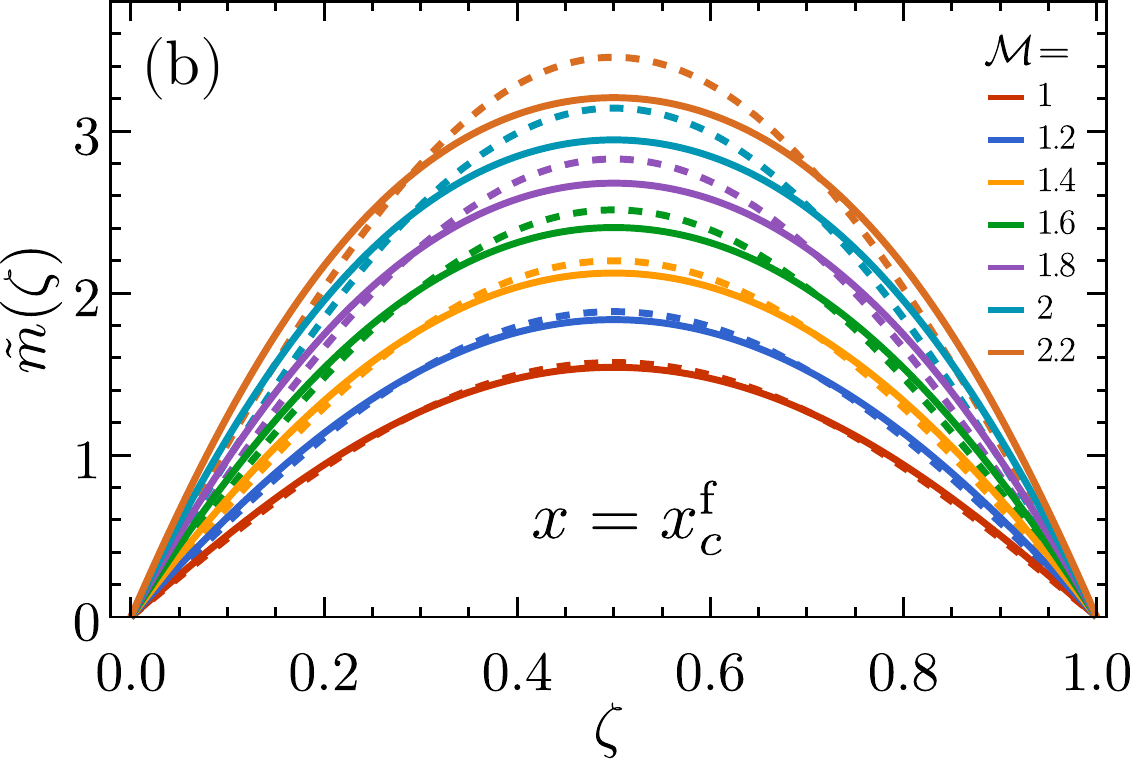}

\caption{The numerical solution $\tilde{m}(\zeta)$ (continuous curves) of the nonlinear MFT in the canonical ensemble is compared with $\tilde{m}_{0}(\zeta)$ (Eq. (\ref{m0tilde}), dashed curves) for $x=0$ ($a$) and $x=x_c\ut{f}$ [see \Eref{eq_x_c_film}] ($b$), for various values of the imposed mass $\mathcal{M}$. }
\label{constrained_profiles}
\end{figure}

Here we consider the constrained profiles obtained by numerically solving the ELE [Eq.~\eqref{eq_ELE}] and compare them with the first-order perturbative solution [\Eref{m0tilde}]. In Fig.~\ref{constrained_profiles}, the two cases $x=0$ and $x=x_{c}^{\textrm f}=-\pi^{2}$ [\Eref{eq_x_c_film}] are examined, where the latter corresponds to the film critical point. It is interesting to note that for $x=x_{c}\ut{f}$ the perturbative profile is not singular, but reduces to a particularly compact form:
\begin{equation}
\lim_{x\rightarrow x_c\ut{f} } \tilde{m}_{0}(\zeta) = \frac{\pi \mathcal{M}}{2} \sin\left(\pi \zeta\right).
\end{equation}

\section{Phase diagrams, equation of state, and scaling}
\label{sec_phasediag}

Here we explore the magnetization phase diagram, the equation of state $\mathcal M (x,B)$, and, in particular, we compare the film behavior with the one corresponding to the bulk. Exact numerical results are discussed in Sec.~\ref{Mexactsec}, while the validity of the perturbative MFT results is studied in Sec.~\ref{Mcomp}. In Sec.~\ref{widom} we show that the near-critical behavior of the mass can be captured by simple scaling arguments. This scaling behavior can even be applied to the order parameter profiles themselves, as will be discussed in Sec.~\ref{WidomProfileSect}.

\subsection{Exact numerical results for the mass}
\label{Mexactsec}

\begin{figure}[t]
\includegraphics[width=.99\columnwidth]{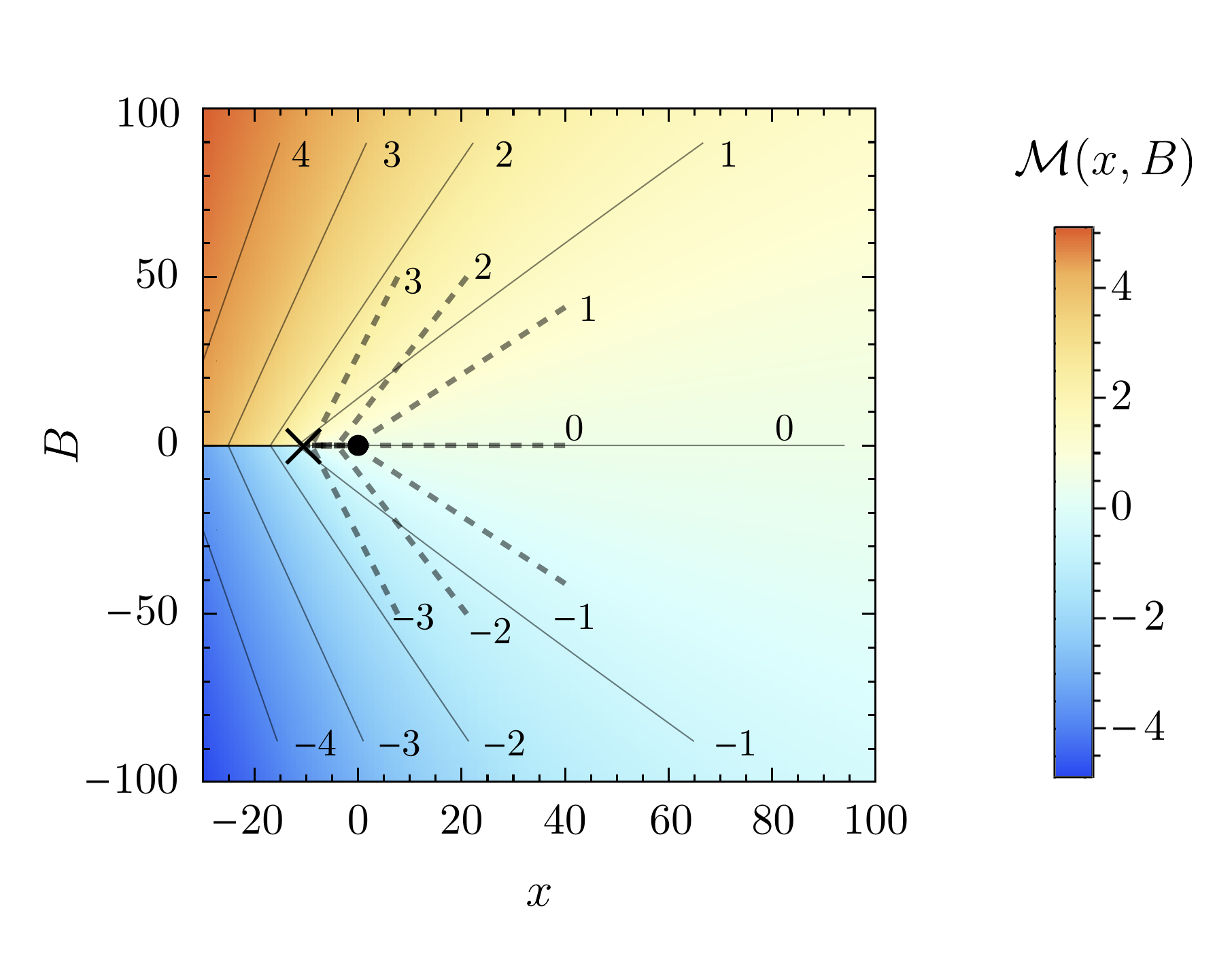} 
\caption{Phase diagram and equation of state of a film with Dirichlet boundary conditions obtained numerically within nonlinear MFT. The color code indicates the value of the mass as a function of the scaled bulk field $B$ and the scaled reduced temperature $x$. Solid lines are iso-$\mathcal M$ lines for the film, while overlayed dashed lines correspond to the bulk system. The cross ($\times$) indicates the film critical point ($x=x_c\ut{f}=-\pi^2$, \Eref{eq_x_c_film}), and the dot ($\bullet$) the bulk critical point ($x=0$).}
\label{phasediagsuperimp}
\end{figure}

While the nonlinear ELE in Eq.~\eref{ELE_MFT}, subject to Dirichlet boundary conditions, can be solved by standard numerical methods for $x>x^{\textrm f}_c$ (i.e., above phase separation in the film) and for sufficiently small $B$ (for which nonlinear effects are not too strong), these methods typically become inaccurate outside these regimes, where gradients of the profile can be large. This issue can be addressed by solving the ELE via the so-called symplectic integration method \cite{ruth_canonical_1983, hairer_geometric_2010,gross_critical_2016}, which, by construction, yields a {spatially} constant pressure in equilibrium. Essentially the ELE in Eq.~\eqref{eq_ELE} is equivalent to the Hamiltonian ``equations of motion'', and the algorithm conserves the Hamiltonian density $\mathcal H=(m')^2/2 - x m^2/2 - m^4/4 + B m$, which, in turn, allows one to directly extract the film pressure $p_f=(\Delta_0/L^4)\mathcal H$ (see, c.f., Eq.~\eqref{eq_pf_landau}). This method has the advantage that it avoids the (inaccurate) numerical computation of $m'$. The order parameter profile obtained this way for a pair ($x,B$) of scaling variables can be integrated numerically in order to determine the corresponding mass. The results of this procedure are shown in Fig.~\ref{phasediagsuperimp}. The shift of the critical point in the film is clearly visible, as is the symmetry $\mathcal M(x,-B) = -\mathcal M(x,B)$. The super-imposed bulk diagram was obtained by solving Eq.~\eref{ELE_MFT}, without the gradient term, for $m=\const$.

\subsection{Comparing exact and perturbative results for the mass}
\label{Mcomp}
\subsubsection{Mass as function of an external field at $x=0$}
The lowest order perturbative MFT solution for the mass [Eq.~\eref{B0const}] is linear in $B$ at bulk $T_c$, i.e., $x=0$. In \fref{fig:M(x=0,B)} we compare this result (solid line) with the exact mass computed from the numerical solution of the nonlinear MFT (dots). The lowest order MFT result starts to deviate significantly from the exact result at $B\gtrsim 10$, whereas the numerical solution gradually approaches the bulk critical behavior $\Mass\propto B^{1/\delta}$ with $\delta=3$ within MFT.
\begin{figure}[t]
\centering
   \includegraphics[width=.99\columnwidth]{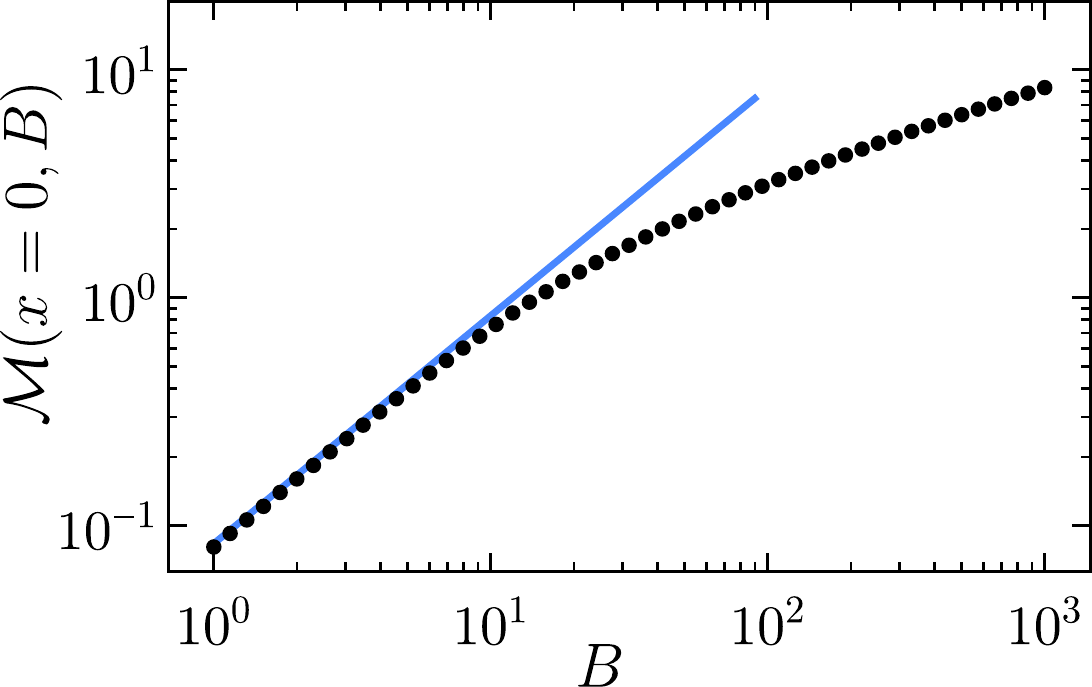}
    \caption{Mass $\Mass$ at the bulk critical point $x=0$ as a function of the scaled bulk field $B$. The solid blue straight line represents the lowest order MFT result [\Eref{B0const}], while the black dots provide the numerical solution of the nonlinear MFT.}
    \label{fig:M(x=0,B)}
\end{figure}

\subsubsection{Mass as function of $x$  with $B=0$}

In the absence of the external magnetic field one can use the exact solution (Eq. (\ref{m_exact})) for the study of the mass:
\begin{equation}
\label{mass_01}
\mathcal{M}(x,B=0) = \int_{0}^{1}d \zeta \, m_{\textrm{exact}}(\zeta,x) \, .
\end{equation}
The elliptic modulus $k=k(x)$ entering into the exact solution is the positive root of the implicit equation $-x=4K^{2}(k^{2})(1+k^{2})$, with $x$ defined in \Eref{eq_x_neg}. The integration in Eq.~(\ref{mass_01}) can be carried out in closed form by using elementary properties of elliptic functions \cite{gradshteyn_table_2014, olver_nist_2010}:
\begin{equation}
\label{mass_02}
\mathcal{M}(x,0) = 2\sqrt{2}\tanh^{-1}\left(k(x)\right) .
\end{equation}
From this result one can easily extract the asymptotic behavior of the mass. In particular, we proceed to analyze Eq. (\ref{mass_02}) for (i) $x$ close to film criticality $T_{\textrm{c}}^{\textrm{f}}$, i.e., for $x\lesssim\xcf=-\pi^{2}$, and for (ii) extreme subcritical temperatures $x\ll -1$.

\underline{(i) $x\lesssim\xcf$:} It is convenient to parametrize the deviation from the critical point as 
\al{
x=-\pi^{2}(1+t^{\textrm f}) \, ,
\label{txrelation}
}
where $\tf=(T_{\textrm{c}}^{\textrm{f}}-T)/(T_{\textrm{c}}^{\textrm{b}}-T_{\textrm{c}}^{\textrm{f}})$ is the film-analogue of the bulk reduced temperature $t$ as introduced in \Eref{eq_tred}. We note that for $T=T_{\textrm{c}}^{\textrm{f}}$ from Eq.\ (\ref{eq_x_neg}) we have {$x=\xcf = (L/\amplXim)^{1/\nu} (T_c^\textrm{f}-T_c^\textrm{b})/T_\textrm{c}^\textrm{b}$}, while instead at $T=T_{\textrm{c}}^{\textrm{b}}$ we recover $x=0$, as expected. {It follows furthermore that $T_\textrm{c}^\textrm{f}(L\to\infty)/T_\textrm{c}^\textrm{b} = 1-\pi^2 (L/\amplXim)^{-1/\nu}$.} We focus on the regime $\tf\rightarrow0^{+}$. Since in this limit $x\rightarrow -(\pi^{2})^{-}$ (see Eq. (\ref{txrelation})), the implicit equation $-x=4K^{2}(k^{2})(1+k^{2})$ can be substituted by its Taylor expansion around the desired value of $k=0$. The corresponding small-modulus expansion of the complete elliptic integral $K(k^{2})$ is $K(k^{2}) = \frac{\pi}{2} \left( 1+ \frac{k^{2}}{4} + \frac{9}{64}k^{4} + \mathcal{O}(k^{6}) \right)$, 
which implies the following expression for $\tf$:
\begin{equation}
\label{mass_04}
\tf = \frac{3}{2} k^{2} + \frac{27}{32}k^{4} + \frac{39}{64}k^{6} + \mathcal{O}(k^{8}).
\end{equation}
The solution $(\tf,k)=(0,0)$, corresponding to $(x,k)=(-\pi^{2},0)$, is trivially reproduced. For $\tf\rightarrow0^{+}$, Eq.~(\ref{mass_04}) gives, to leading order, $k = \sqrt{\frac{2}{3}\tf}$,
which is valid for $x$ approaching $-\pi^{2}$ from below. Higher order corrections 
can be obtained by iterating this procedure. Inserting this result into Eq.~(\ref{mass_01}) and using the fact that $\tanh^{-1}(k\rightarrow0)=k+\mathcal{O}(k^{3})$, one obtains the following scaling behavior:
\begin{equation}
\label{mass_07}
{\mathcal{M}(x,0) \simeq \frac{4}{\sqrt{3}} (\tf)^{\beta} }
, \quad \tf\rightarrow 0^{+} \;\,\textrm{or} \;\,x \rightarrow \left(-\pi^{2}\right)^{-} ,
\end{equation}
with the exponent $\beta=\frac{1}{2}$. Equation \eqref{mass_07} is valid in the asymptotic regime $\tf\rightarrow 0^{+}$ where successive corrections $~\sim (\tf)^{\hat{\beta}}$, characterized by an exponent $\hat{\beta}>\beta$, vanish faster than $(\tf)^{\beta}$.

\begin{figure}[t]
\centering
   \includegraphics[width=.99\columnwidth]{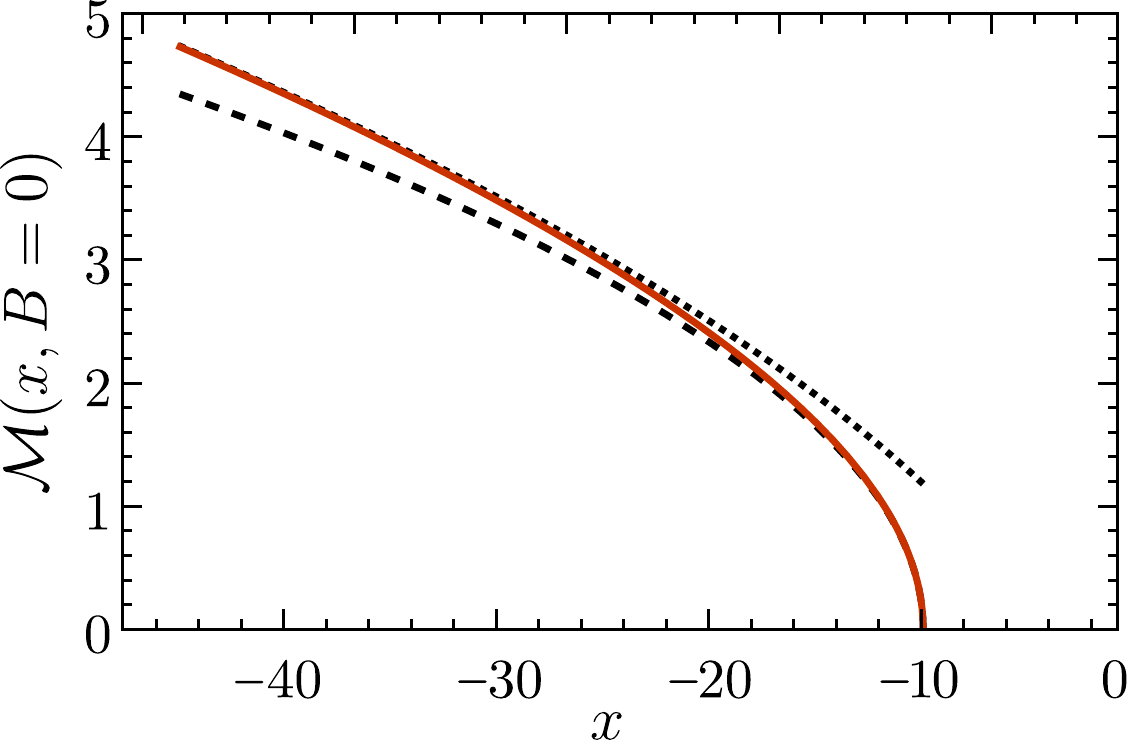}
    \caption{Mass $\Mcal$ in a film with Dirichlet boundary conditions for $B=0$ as a function of $x$, as given by the exact MFT expression in \Eref{mass_02} (red solid curve). The asymptotic behavior described in \Eref{mass_07} (dashed curve) and in \Eref{mass_10} (dotted curve) agree with the exact expression for $x=x_c\ut{f}\approx - 9.87$ and for large values of $-x$, respectively.}
    \label{fig:M(x,B=0)}
\end{figure}

\underline{(ii) $x\ll-1$:} Since $-x \gg1$, the roots of $x(k)$ accumulate towards $k=1^{-}$. Writing $k=1-\epsilon$ for certain $\epsilon\rightarrow 0^{+}$, one has $\tanh^{-1}(1-\epsilon)=\frac{1}{2}\ln\frac{2}{\epsilon}+ \frac{1}{2}p_{n}(\epsilon) + \mathcal{O}(\epsilon^{n+1}) $, where $p_{n}(\epsilon)$ is a polynomial in $\epsilon$ of degree $n$. Hence for large $|-x|$ the mass is approximately given by $\mathcal{M}(x,0) \simeq \sqrt{2}\ln\frac{2}{\epsilon} + \sqrt{2}p(\epsilon) $. 
In order to identify the small parameter $\epsilon$ in terms of $x$, we note that if the elliptic modulus approaches unity, $K(k^{2}) = -\frac{1}{2}\ln\left(\frac{1-k^{2}}{16}\right) \bigl[ 1+ \mathcal{O}(1-k)\bigr]$, 
so that the implicit equation for $k(x)$ exhibits the asymptotic behavior $k(x) \simeq 1-8 \, \textrm{e}^{-\sqrt{-x/2}} = 1-\epsilon$. Accordingly, the mass is $\mathcal{M}(x,0) = \sqrt{-x} - 2\sqrt{2}\ln2 + \sqrt{2}p(\epsilon)$, where the last term is negligible because $\epsilon\rightarrow0$ and $p(0)=0$. This renders the asymptotic result
\begin{equation}
\label{mass_10}
{\mathcal{M}(x,0) \simeq \sqrt{-x} - 2\sqrt{2}\ln2}, \quad x\rightarrow -\infty .
\end{equation}

To summarize, we have derived the analytical expression of the mass $\mathcal{M}$ in the absence of an external field, and its asymptotic behavior close to film criticality ($x \lesssim \xcf=-\pi^{2}$) and far from criticality in the two-phase region ($x \ll -1$). As shown in Fig. \ref{fig:M(x,B=0)}, the approximate expressions agree well with the analytical result in \Eref{mass_02}.

\subsection{Widom scaling for the mass}
\label{widom}
It is well known \cite{gambassi_critical_2006,nakanishi_critical_1983} and explicitly demonstrated in Sec.~\ref{B0solution}, that in the film geometry the presence of two confining walls induces a shift of the bulk critical point from $x_c^{\textrm{b}}=0$ to $\xcf = -\pi^{2}$. 
In the present section we discuss in detail the mean-field critical behavior around $\xcf$, resulting from \Eref{ELE_MFT}.

\begin{figure}[t]
\centering
\includegraphics[width=.99\columnwidth]{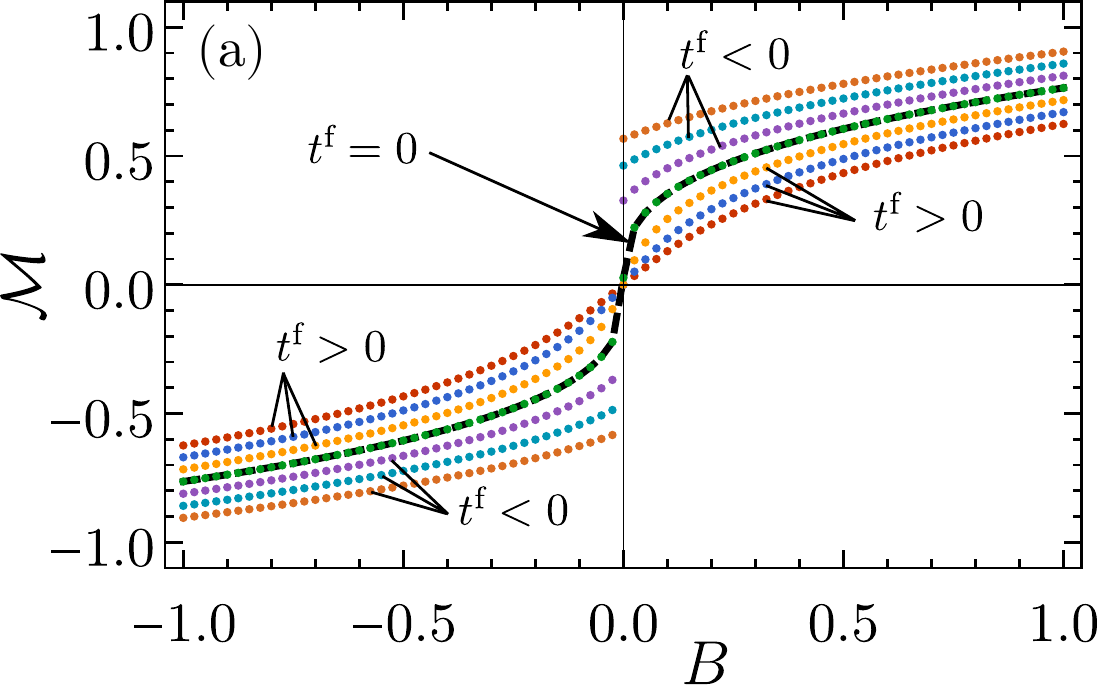}

\vspace{.2cm}
\includegraphics[width=.99\columnwidth]{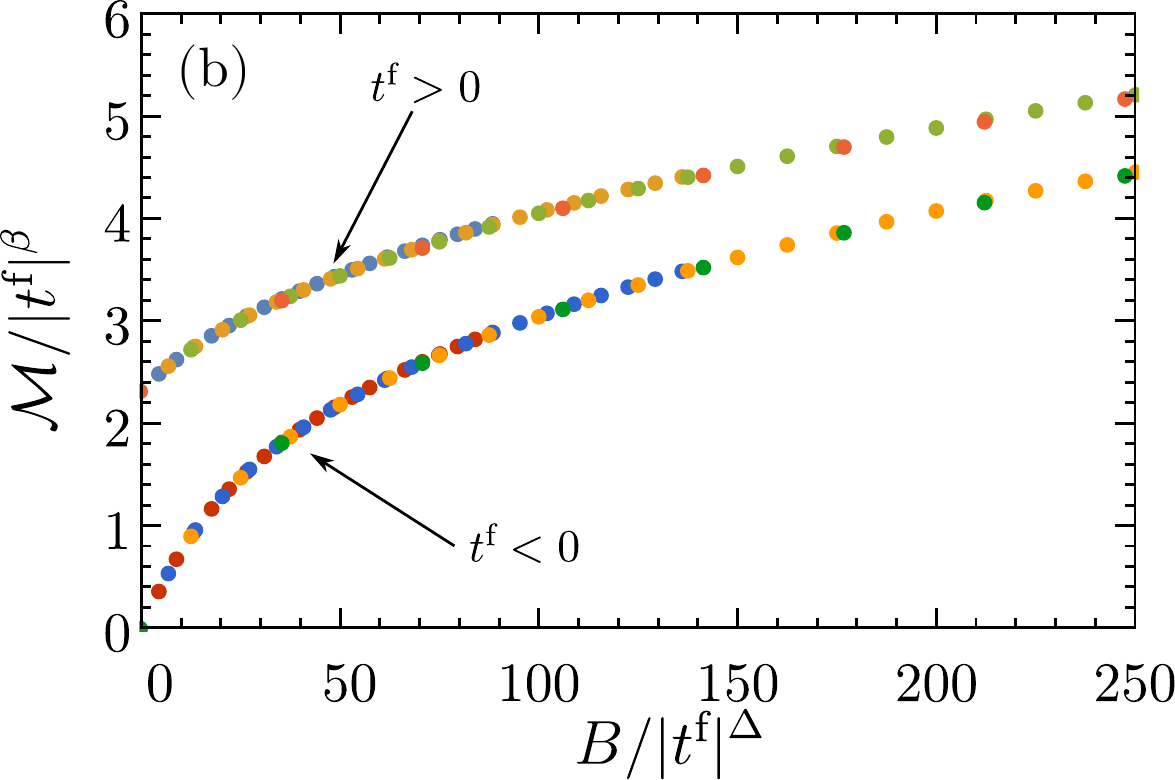}
\caption{(a) A section of the equation of state around the film critical point in the phase diagram. The critical isotherm corresponding to $\tf=0$ [see Eq. (\ref{critical02})] is shown by the dashed black curve, while symbols represent numerical data. (b) Test of the scaling hypothesis [\Eref{scaling_region}] for data contained in panel (a). The points corresponding to the ``supercritical'' regime ($\tf<0$) collapse onto the scaling function $\mathcal{U}_{-}$, while data for the ``subcritical'' regime ($\tf>0$) collapse onto $\mathcal{U}_{+}$.}
\label{WidomFig}
\end{figure}

It is useful to recall the essential ideas of the static scaling hypothesis, as originally formulated by Widom \cite{Widom_equation_1965, stanley_introduction_1971}. The film critical point is located at $(\tf,B)=(0,0)$, where $\tf$ [see \Eref{txrelation}] is the reduced temperature of the film relative to $T_{\textrm{c}}^{\textrm{f}}$. Instead of considering the order parameter profile inside the film, here we are interested in the mass $\mathcal{M}(\tf,B) = \mathcal{M}(x(\tf),B)$. In the critical region of the film, for a vanishing bulk field $B$ one expects the scaling behavior 
\al{
\mathcal{M}(\tf,B=0) =
\begin{cases}
 0,\quad& \tf<0\;\; \\
\pm \mathcal{C}_{\tf} |\tf|^{\beta} ,\quad& \tf > 0 \;.
\end{cases}
\label{critical01}
}
We note that, according to Eq. (\ref{txrelation}), $t^{\textrm{f}}<0$ $[t^{\textrm{f}}>0]$ corresponds to $x>\xcf$ $[x<\xcf]$. The critical isotherm follows as 
\begin{equation}
\label{critical02}
\mathcal{M}(\tf=0,B) = \mathcal{C}_{B} \, \textrm{sign}(B) \, |B|^{1/\delta}.
\end{equation}
The above relations can be considered as a definition of the critical exponents $\beta$ and $\delta$ and of the non-universal amplitudes $\mathcal{C}_t$ and $\mathcal{C}_B$. According to the scaling hypothesis, in the near-critical region around $T_{\textrm{c}}^{\textrm{f}}$ the equation of state fulfills a homogeneity relation of the form 
\al{
\mathcal{M}(\tf,B) =
\begin{cases}
 (-\tf)^{\beta} \mathcal{U}_{-}\left(B/(-\tf)^{\Delta}\right) ,\quad & \tf<0,\\
 (\tf)^{\beta} \mathcal{U}_{+}\left(B/(\tf)^{\Delta}\right) ,\quad & \tf>0, 
\end{cases}
\label{scaling_region}
}
where $\mathcal{U}_{\pm}$ are a pair of universal scaling functions and $\Delta=\beta\delta$  is called the \emph{gap exponent} \cite{Kadanoff_static_1967, stanley_introduction_1971}. 
Various sections of the phase diagram in the scaling region lead to curves of the type shown in Fig.~\ref{WidomFig}(a). A suitable rescaling of the thermodynamic variables $t$ and $B$ as prescribed by Eqs.~\eqref{critical01} and  \eqref{critical02} results in a data collapse onto two single master curves corresponding to the scaling functions $\mathcal{U}_{\pm}$ (see Fig. \ref{WidomFig}(b)).
In the previous subsection we have established \Eref{critical01} in the form of \Eref{mass_07}, leading to $\beta=\frac{1}{2}$ and to the non-universal amplitude $\mathcal{C}_{\tf} = \frac{4}{\sqrt{3}}$.
On the other hand, the results of the complete numerical analysis, shown in Fig.~\ref{WidomFig}, confirm \Eref{critical02}; in fact the critical isotherm in the scaling region near $\xcf$ can be approximated well by \Eref{critical02} with the critical exponent $\delta=3$ and the non-universal amplitude
$\mathcal{C}_{B} \simeq 0.76.$ 
We thus recover $\Delta=\frac{3}{2}$ for the gap exponent and obtain an excellent data collapse. 

To summarize, our analytical and numerical analysis recovers the expected mean field critical exponents for the film critical point. 
We remark that the maximum value of the critical profile (in the center of the film) has the same scaling behavior as the total mass, $m(\zeta=\frac 1 2;\xcf,B)=\widehat{\mathcal{C}}_{B} \, \textrm{sign}(B) \, |B|^{1/\delta}$, with $\widehat{\mathcal{C}}_{B} \simeq 1.19$ and $\delta=3$. 

This analysis reveals explicitly that, as expected, within MFT the bulk transition in spatial dimension $d$ exhibits the same scaling behavior and the same critical exponents as its counterpart in the film which, asymptotically, behaves as an effectively $(d-1)$-dimensional system. The inability to capture this actual dimensional crossover is a well-known shortcoming of {many} analytical approaches, i.e., MFT and beyond \cite{krech_casimir_1994, diehl_theory_1997, dohm_critical_2009, diehl_large-$n$_2014} (see, however, Refs.\ \cite{kastening_finite-size_2010, dohm_crossover_2018}), whereas simulations can deal with this issue successfully.

\subsection{Magnetization profiles in the near-critical region: insights from Widom scaling}
\label{WidomProfileSect}

Here we consider the case $B=0$ and $x\lesssim x_{c}$. Since at criticality the profile vanishes, the ELE in Eq.~\eqref{ELE_MFT} in the vicinity of the film critical point, i.e.,
\al{
m''(\z) - \xcf m(\z)  - m^3(\z) = 0,
\label{ELE_MFT_criticality_1}
}
can be approximated by the linearized equation 
\al{
m''(\z) + \pi^{2} m(\z) = 0 ,
\label{ELE_MFT_criticality_2}
}
because the cubic term is smaller that the linear terms. Equation \eqref{ELE_MFT_criticality_2} with Dirichlet boundary conditions is solved by
\begin{equation}
\label{ }
m_{\textrm{lin}}(\zeta) = \mathcal{A} \sin(\pi\zeta).
\end{equation}
However, the amplitude $\mathcal{A}$ cannot be fixed by Eq. (\ref{ELE_MFT_criticality_2}), because the cubic term has been neglected. Nonetheless, we can determine $\mathcal{A}$ by considering a suitable limit of the exact solution. For $x\rightarrow -\pi^{2}$ we can use the reduced temperature $\tf \rightarrow 0$ from \Eref{txrelation}. We recall that within this limit the elliptic modulus is $k = \sqrt{\frac{2}{3}\tf}$ and that for vanishing $k^{2}$ the Jacobi elliptic function $\textrm{sn}(w;k^{2}\rightarrow0) \rightarrow \sin(w)$ reduces to a standard sine function. Therefore in the limit $\tf\rightarrow0^{+}$ \Eref{m_exact} produces exactly
\begin{equation}
\label{scaling_prof_t}
m(\zeta,x \lesssim \xcf,B=0) = \frac{2\pi \sqrt{\tf}}{3} \sin(\pi\zeta).
\end{equation}
Spatial integration yields the mass $\mathcal{M}(x,B=0)= 4 \left(\frac{\tf}{3}\right)^{1/2}$ for $x\rightarrow \xcf$.
The behavior of the scaled bulk field $B$ is less obvious. In the previous section we noted a scaling behavior for the maximum value of the magnetization profiles, namely $m(\zeta=1/2,B) \sim B^{1/3}$. The same behavior extends, with remarkably good agreement with the numerical results of Fig. \ref{uncontrained_profiles}, also to $\zeta\neq 1/2$. 
We find that \Eref{scaling_prof_t} follows an analogous scaling, i.e.,
\begin{equation}
\label{scaling_prof_B}
m(\zeta,x=\xcf,B \simeq 0) = \widehat{\mathcal{C}}_{B} B^{1/3} \sin(\pi\zeta) .
\end{equation}
Combinining \Eref{scaling_prof_t} and \Eref{scaling_prof_B}, in the scaling region around the film critical point we have
\begin{equation}
\label{51}
m(\zeta,x \lesssim \xcf,B \simeq 0) = (\tf)^{\beta}\Phi\left(\frac{B}{(\tf)^{\Delta}}\right) \sin(\pi\zeta) ,
\end{equation}
with a scaling function $\Phi$. Since $\int_{0}^{1}d \zeta \, \sin(\pi\zeta)=\frac{2}{\pi}$, one has
\begin{equation}
\mathcal{M}(x \simeq x_{c},B \simeq 0) = (\tf)^{\beta} \Psi\left(\frac{B}{(\tf)^{\Delta}}\right) ,
\end{equation}
where $\Psi(u)=(2/\pi)\Phi(u)$, so that \Eref{scaling_region} is recovered, for which we identify $\Psi(u)$ as $\mathcal{U}_{+}(u)$. Thus, the scaling functions computed for the mass equation of state in Fig.~\ref{WidomFig} capture well the spatially integrated order parameter profiles in the near-critical region.

\section{Critical Casimir Force}
\label{sec_Forces}

In this section we study the critical Casimir force (CCF) in the grand canonical and the canonical ensemble, for a film subject to Dirichlet boundary conditions. We therefore briefly recall the general definitions and protocols for computing the CCF, as set out in Sec.~III of Ref.~\cite{gross_critical_2016}.

In general, the equilibrium CCF $\mathcal K$ provides the derivative of the residual free energy, or, in terms of the stress tensor, quantifies the change of the free energy of the film upon shifting the position of the boundaries. In the first case, one has
\al{
\Kcal = -\frac{d\Fcal\res}{dL},
\label{eq_pCas_dFdL}
}
where we have decomposed the free energy of the film according to
\al{
\Fcal_f = L (-p_b) + f_s + \Fcal\res,
\label{eq_Ffilm_split}
}
in terms of the bulk pressure $p_b$, the surface free energy $f_s$, and the residual free energy $\Fcal\res$ (all per transverse area $A$ and $k_B T$). 
Generally, the bulk term scales $\propto L$ and the surface term $\propto L^0$, while the residual terms vanish exponentially for $L\to\infty$ (see, e.g., Refs.\ \cite{gross_critical_2016, privman_finite-size_1990}).
We remark that for realistic fluid films, long-ranged van der Waals forces provide algebraically decaying non-universal contributions to the residual free energy \cite{dantchev_universality_2003,dantchev_excess_2006,dantchev_interplay_2007,dohm_diversity_2008}. In the present study, we consider only the universal critical Casimir contribution.

\begin{widetext}
In the second case, the CCF is the difference between the film pressure $p_{f} = -d\Fcal_{f}/dL$ and the pressure of the surrounding bulk medium in which the film is immersed: 
\al{
\Kcal = p_f - p_b.
\label{eq_pCas_pdiff}
}
The bulk pressure is naturally defined as 
\al{
p_b = \lim_{L\to\infty} p_f,
\label{eq_pB_lim_pF}
}
where the limit is performed by keeping fixed the relevant thermodynamic control parameters (i.e., the chemical potential $\mu$ for the grand canonical ensemble, and the mass density $\mden = \mass/L$ for the canonical ensemble) \footnote{In the canonical ensemble the actual thermodynamic control parameter is the total mass $\mass$. However, as discussed in detail in Ref.\ \cite{gross_critical_2016}, for the purpose of determining the finite-size limit in \Eref{eq_pB_lim_pF}, instead the mass density $\mden$ should be kept fixed.}.

The CCF (per transverse area $A$ and $k_B T$) in the \textit{g}rand \textit{c}anonical and the \textit{c}anonical ensemble takes the following scaling form \cite{krech_casimir_1994,brankov_theory_2000,gross_critical_2016}:
\begin{subequations} \begin{align}
\Kcal\gc(t,\mu,L) &= L^{-d} \Xi\gc\left(  \left(\frac{L}{\amplXip}\right)^{1/\nu}t, \left(\frac{L}{\amplXimu}\right)^{\Delta/\nu}\mu \right), \label{eq_Casi_force_gc} \\
\Kcal\can(t,\mden,L) &= L^{-d} \Xi\can\left(  \left(\frac{L}{\amplXip}\right)^{1/\nu}t, \left(\frac{L}{\amplXip}\right)^{\beta/\nu} \frac{\mden}{\amplPhit} \right), 
\label{eq_Casi_force_can}
\end{align}\label{eq_Casi_force}\end{subequations}
where $\Xi\gc$ and $\Xi\can$ are scaling functions, which will be determined below for $t=(T-T_{\textrm{c}}^{\textrm{b}})/T_{\textrm{c}}^{\textrm{b}}>0$ and within MFT, whereby we take the values of the critical exponents pertaining to $d\geq 4$ spatial dimensions. 
The scaling relation in \Eref{eq_Casi_force} expresses the two-scale factor universality \cite{privman_universal_1984, privman_finite-size_1990} valid for simple fluids below the upper critical dimension $d=4$. Within MFT, the scaling functions $\Xi\cgc$ acquire an \emph{a priori} undetermined prefactor $\Delta_0$ involving the coupling constant $g$ [\Eref{eq_Delta0}]. Accordingly, we shall present our results within MFT in terms of \emph{reduced} scaling functions $\Xi\cgc/\Delta_0$.
\end{widetext}

Instead of using \Eref{eq_pB_lim_pF}, the film pressure can equivalently be obtained from the stress tensor $T_{ij}$:
\al{
p_f = T_{zz}[\phi\eq] = -\frac{d }{d L}\Fcal_f[\phi\eq],
\label{eq_stressten_dFdL}
}
where $T_{zz}[\phi\eq]$ is computed from the order parameter profile minimizing $\mathcal F_f$ (Eqs. (\ref{eq_Landau_func_c}) and (\ref{eq_Landau_func_gc})). Note that here we have assumed the boundaries of the film to be normal to the $z$-direction. Analogously, the bulk pressure can be obtained from the corresponding bulk order parameter at equilibrium, $p_b=T_{zz}(\phi_{b})$. Therefore \Eref{eq_pCas_pdiff} allows one to compute $\Kcal$ without explicitly evaluating derivatives of free energy functionals. In the grand canonical ensemble, the definitions of $\Kcal$ in Eqs.~\eref{eq_pCas_dFdL} and \eref{eq_pCas_pdiff} yield equivalent results, whereas differences may appear due to additional surface contributions in the canonical ensemble \cite{gross_critical_2016}. We recall that, in thermal equilibrium, $T_{zz}[\phi\eq]$ is in general independent of $z$.

A core result of Ref.~\cite{gross_critical_2016} is that the stress tensor in the canonical ensemble can be computed using a grand canonical stress tensor in which the chemical potential takes the value $\mu = \tilde \mu (\Phi)$, satisfying the mass constraint in \Eref{eq_Mass0},
\al{
T_{ij}\can[\phi\eq] = T_{ij}\gc([\phi\eq];\mu=\tilde \mu),
\label{eq_stressten_can}
}
in terms of the solution $\phi\eq$ of the ELE. By construction, this yields equal film pressures in the two ensembles, $p_f\can[\phi\eq] = p_f\gc([\phi\eq];\tilde \mu) $. In the grand canonical ensemble, the mean field stress tensor corresponding to the free energy functional in \Eref{eq_Landau_func_gc} is \cite{krech_casimir_1994}
\al{
&T_{ij}^{\textrm{(gc)}}([\phi_{eq}];\mu) = (\partial_i\phi_{eq})(\partial_j\phi_{eq})\nl
&- \delta_{ij}\big[\frac 1 2 \sum_k (\partial_k\phi_{eq})(\partial_k\phi_{eq})  
+ \frac 1 2\tau \phi_{eq}^2 + \frac{1}{4!}g\phi_{eq}^4-\mu\phi_{eq}
\big],
}
giving rise to the film pressure
\beq 
\begin{split}
p_f^{\text{(c,gc)}} &= T_{zz}^{\text{(c,gc)}} =  \onehalf (\pd_z \phi\eq)^2  - \onehalf \tau \phi\eq^2 - \frac{1}{4!} g \phi\eq^4 + \tilde \mu \phi\eq  \\
&=\frac{\Delta_0}{L^{4}} \left[ \onehalf \left(\pd_\zeta m\eq\right)^2  - \onehalf \tscal m\eq^2 - \frac{1}{4} m\eq^4 + \tilde B m\eq  \right],
\end{split}
\label{eq_pf_landau}
\eeq
where the dimensionless variables from \Eref{eq_scalvar_mft} have been re-introduced; $\Delta_{0}$ is given by Eq. (\ref{eq_Delta0}). In turn, the bulk pressure in the grand canonical ensemble,
\al{
p_b\gc(\mu_b\gc) =  \onehalf \tau\phi^2_b + \frac{1}{8}g\phi^4_b,
\label{eq_pB_gc}
}
is obtained by solving the bulk equation of state (i.e., the ELE without gradient terms),
\al{
\tau\phi_b + \frac{1}{6}g\phi_b^3 = \mu_b\gc =\mu\,,
\label{eq_bulk_EOS_gc}
}
in order to find the spatially constant solution $\phi_b$, and to insert it into \Eref{eq_stressten_can}. By virtue of the grand canonical coupling between film and bulk, the chemical potential $\mu$ here is the same as for the film. 

In contrast, in the canonical ensemble, the film and the bulk system are constrained to have the same mass density $\varphi$, which gives rise to the following canonical bulk pressure:
\begin{widetext}
\beq p_b\can(\phi_b) = 
 \onehalf \tau \phi_{b}^2 + \frac{1}{8}g\phi_{b}^4 ,\qquad \text{with}\quad  
 \begin{cases}
   \phi_b = \pm \phi_{b,\text{eq}},\qquad &\tau<0\quad \text{and}\quad -\phi_{b,\text{eq}} \leq \varphi \leq \phi_{b,\text{eq}},\\
   \phi_b = \varphi,\qquad &\text{otherwise,}
  \end{cases}
\label{eq_pB_can}
\eeq
where $\phi_{b,\text{eq}}$ denotes the OP minimizing the LG functional in \Eref{eq_Landau_func_c}.
The chemical potential corresponding to the bulk system of mass density $\mden$ is
\beq 
\mu_b\can = \begin{cases}
 0,\qquad &\tau<0\quad \text{and}\quad -\phi_{b,\text{eq}} \leq \varphi \leq \phi_{b,\text{eq}},\\
 \tau\varphi + \frac{1}{6}g\varphi^3,\qquad &\text{otherwise.}
 \end{cases}
\label{eq_bulk_EOS_can}
\eeq 
\end{widetext}
In what follows, we shall focus on the region $\tau>0$, i.e., we avoid bulk phase separation, so that the bulk pressure can be directly obtained as $p_b\can(\phi_b) =  \onehalf \tau \varphi^2 + \frac{1}{8}g\varphi^4$.

As stated, the film pressures are equal in the grand canonical and the canonical ensembles. However, due to the different thermodynamic coupling of film and bulk outlined above, the CCF $\Kcal$ can differ in the respective ensembles. Indeed, this has been reported in Ref. \cite{gross_critical_2016} for the case of critical adsorption whereas here we investigate the CCFs for films with Dirichlet boundary conditions. We proceed by using the linear MFT results from Sec.~\ref{MFTeq} in order to compute the CCF using the stress tensor in \Sref{sec_Forces_ST}. These perturbative expressions are compared with exact numerical MFT results for the canonical and grand canonical scaling functions of the CCF. In Sec.~\ref{sec_Forces_FE} the CCF is computed directly by differentiating the free energy functionals, which are expressed in terms of the perturbatively computed OP profiles.

\subsection{CCF within linear MFT deduced from the stress tensor}
\label{sec_Forces_ST}

We employ the stress tensor in \Eref{eq_pf_landau} in order to compute the film pressure from the order parameter profiles determined in Sec.~\ref{MFTeq}. In the grand canonical case, $T_{zz}$ is determined in terms of the unconstrained OP in the presence of the external field, i.e., for fixed $\mu$. The canonical pressure can be obtained analogously by using the constrained profile $\bar \phi$, where now $\tilde \mu(\varphi=\Phi/L)$ is the {constraint-induced chemical} potential guaranteeing a certain mass density $\varphi$. Rewriting \Eref{eq_pf_landau} as 
\al{
p_f^{\textrm{(c,gc)}} =
\frac{\Delta_0}{L^4} \Tcal
\label{pf}
}
and inserting the expansion of $m$ in terms of powers of $\epsilon$ as defined in Eqs. (\ref{ELE_MFT}) and (\ref{epsexpand}), we find the lowest orders of $\Tcal=\Tcal_0 + \epsilon \Tcal_1 +\ldots$:
\al{
\Tcal_0 &= \frac 1 2 (\partial_\zeta \tilde m_{0})^2 -\frac{\tilde m_0^2 x}{2}
+ \tilde B_0 \tilde m_0, \nl
\Tcal_1 &= (\partial_\zeta \tilde m_{0})(\partial_\zeta \tilde m_{1}) - \tilde m_0 \tilde m_1 x  - \tilde m_0^3 \tilde m_1   +   \tilde B_0 \tilde m_1 + \tilde B_1 \tilde m_0 .
\label{Texpand}
}
At lowest order we have implicitly neglected the $\phi^4$ term in the free energy [and thus also the quartic term in \Eref{eq_pf_landau}], which explains the absence of this term in the expression for $\Tcal_0$.

\subsubsection{Grand canonical CCF}

Using $B$ instead of $B_0$ and inserting the linear MFT solution from Eq.~\eref{m0} into Eqs.~\eref{pf} and \eref{Texpand}, we find
\al{
\Tcal_0^{\textrm{(gc)}} = \frac{B^2 \tanh ^2\left(\frac{\sqrt{x}}{2}\right)}{2 x}.
}
Upon rescaling to dimensional variables via \Eref{eq_scalvar_mft}, we identify the corresponding film pressure
\al{
(p_f^{\textrm{(gc)}})_0 =\frac{\mu ^2 }{2 \tau } \tanh ^2\left(\frac{L \sqrt{\tau }}{2}\right).
\label{pfgclinmft}
}
The corresponding bulk limit, taken with $\mu$ fixed, is
\al{
(p_b^{\textrm{(gc)}})_0 = \frac{\mu ^2}{2 \tau }.
\label{pbc0}
}
The CCF can now be computed by using \Eref{eq_pCas_pdiff}:
\al{
\mathcal K^{\textrm{(gc)}}_0 =  -\frac{\mu ^2}{\tau[1+  \cosh \left(L \sqrt{\tau }\right)]}.
\label{Kgc}
}
From Eq.~\eref{B0const}, the chemical potential corresponding to the mass constraint follows as
\al{
\tilde \mu = \frac{\tau  \varphi }{1-\frac{2}{L \sqrt{\tau }} \tanh \left(\frac{L \sqrt{\tau}}{2}\right)},
\label{eq_mu_constr}}
which, together with \Eref{Kgc}, gives
\al{
\mathcal K^{\textrm{(gc)}}_0 = -\frac{\tau  \varphi ^2}{\left(1+  \cosh \left(L \sqrt{\tau }\right) \right) \left(1-\frac{2 \tanh \left(L \sqrt{\tau}/2\right)}{L \sqrt{\tau }}\right)^2}.
}
According to \Eref{eq_Casi_force_gc} (with $d=4$), the reduced scaling function of the CCF results as
\al{
\frac{\Xi^{\textrm{(gc)}}}{\Delta_0}= 
-\frac{\mathcal M^2 x^2}{\left(1+ \cosh \left(\sqrt{x}\right)\right) \left(\sqrt{x}-2 \tanh \left(\frac{\sqrt{x}}{2}\right)\right)^2}.
\label{Xigc}
}
Note that this scaling function diverges at bulk criticality:
\al{
\frac{\Xi^{\textrm{(gc)}}(x\to 0)}{\Delta_0} \simeq -\frac{72\mathcal M^2}{x}.
\label{Xigcx0}
}
This divergence is entirely due to the bulk pressure in Eq.~\eref{pbc0} and can be considered as an artifact of linear MFT. (An analogous divergence occurs in the case of critical adsorption, see Ref.\ \cite{gross_critical_2016}.)
Far above $T_c$, the scaling function vanishes as
\al{
\frac{\Xi^{\textrm{(gc)}}(x\gg 0)}{\Delta_0} \simeq 2 \mathcal M^2 e^{-\sqrt{x}} x,
\label{Xigcxinf}
}
which is intuitively expected, because the CCF is expected to vanish in the limit of thick films, i.e., $x=(L/\xi)^{1/\nu}\rightarrow\infty$.

\subsubsection{Canonical CCF}

In this case, the constrained linear mean-field profile from Eq.~\ref{m0tilde} yields
\al{
\Tcal_0^{\textrm{(c)}} = \frac{\mathcal M^2 x^2 \tanh ^2\left(\frac{\sqrt{x}}{2}\right)}{2 \left(\sqrt{x}-2 \tanh \left(\frac{\sqrt{x}}{2}\right)\right)^2},
}
which renders the corresponding film pressure (expressed in terms of dimensional variables, see \Eref{eq_scalvar_mft})
\al{
(p_f^{\textrm{(c)}})_0 =\frac{\tau \varphi ^2}{2} \frac{ \tanh ^2\left(\frac{L \sqrt{\tau }}{2}\right)}{ \left(1-\frac{2 \tanh \left(L \sqrt{\tau}/2\right)}{L\sqrt{\tau }}\right)^2}.
\label{pfcstresstens}
}
The same expression results upon inserting \Eref{eq_mu_constr} into \Eref{pfgclinmft}.
In the canonical ensemble, the bulk limit of \Eref{pfcstresstens} is obtained by keeping a fixed mass density $\varphi$ [see \Eref{eq_pB_lim_pF}], 
\al{
(p_b^{\textrm{(c)}})_0 = \frac{\tau  \varphi ^2}{2}.
\label{pbc}
}
Subtracting \Eref{pbc} from \Eref{pfcstresstens} leads to the CCF in the canonical ensemble:
\al{
\mathcal K^{\textrm{(c)}}_0 =& \frac{\tau \varphi ^2}{2} \left[\frac{ \tanh ^2\left(\frac{L \sqrt{\tau }}{2}\right)}{ \left(1-\frac{2 \tanh \left(L\sqrt{\tau}/2\right)}{L\sqrt{\tau }}\right)^2}- 1\right],
\label{Kc}
}
which can be brought into the scaling form given in \Eref{eq_Casi_force_can} with the reduced scaling function
\al{ 
\frac{\Xi^{\textrm{(c)}}}{\Delta_0}&= \frac{\mathcal M^2 x}{2} \Bigg[\frac{\tanh^2\left(\frac{\sqrt x}{2}\right)}{\left(1-\frac 2 {\sqrt{x}} \tanh \left(\frac{\sqrt{x}}{2}\right)\right)^2}-1 \Bigg].
\label{Xican}}
A comparison with the grand canonical CCF from \Eref{Xigc} reveals that 
\al{
\frac{\Xi^{\textrm{(c)}}}{\Delta_0}&=\frac{\Xi^{\textrm{(gc)}}}{\Delta_0} 
+ \frac{\mathcal M^2 x}{2} \Bigg[\frac{\tanh^2\left(\frac{\sqrt x}{2}\right)}{\left(1-\frac 2 {\sqrt{x}}\tanh \left(\frac{\sqrt{x}}{2}\right)\right)^2}-1\Bigg]\nl
&= -\frac{\mathcal M^2 x^2}{\left(1+ \cosh \left(\sqrt{x}\right)\right) \left(\sqrt{x}-2 \tanh \left(\frac{\sqrt{x}}{2}\right)\right)^2} \nl
&\quad+\frac{\mathcal M^2 x}{2} \Bigg[\frac{\tanh^2\left(\frac{\sqrt x}{2}\right)}{\left(1-\frac 2 {\sqrt{x}} \tanh \left(\frac{\sqrt{x}}{2}\right)\right)^2}-1 \Bigg].
\label{Xic_gc_comp}
}
Different from the grand canonical scaling function [\Eref{Xigc}], the canonical one attains a finite value at bulk criticality:
\al{
\frac{\Xi^{\textrm{(c)}}(x\to 0)}{\Delta_0} \simeq 18 \mathcal M^2.
\label{Xicx0}
}
However, for thick films ($x=(L/\xi)^{1/\nu}\rightarrow\infty$) the canonical scaling function diverges as
\al{
\frac{\Xi^{\textrm{(c)}}(x\gg 1)}{\Delta_0} \simeq 2 \mathcal M^2 \sqrt{x}.
\label{Xicxinf}
}
This divergence is essentially a consequence of the OP constraint, as can be seen by inserting the {constraint-induced chemical} potential $\tilde \mu$ [\Eref{eq_mu_constr}] into \Eref{pfgclinmft} in order to yield the canonical film pressure in \Eref{pfcstresstens}. 
We demonstrate below  [see, c.f., \Eref{eq_CCF_c_Fres}] that the divergence stems from a \emph{surface} contribution to the canonical film pressure.

\subsection{Discussion of the CCF obtained within linear MFT and comparison with full, numerical results}

\begin{figure}[t]
\includegraphics[width=.99\columnwidth]{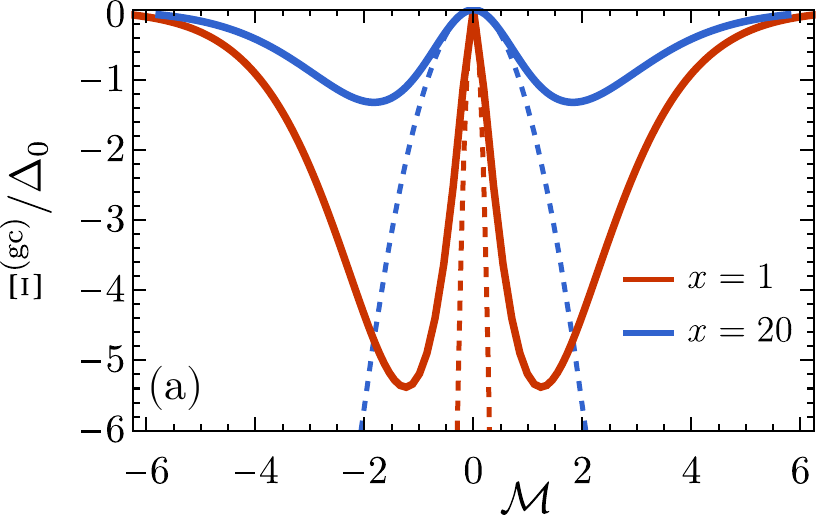} 

\vspace{0.3cm}
\includegraphics[width=.99\columnwidth]{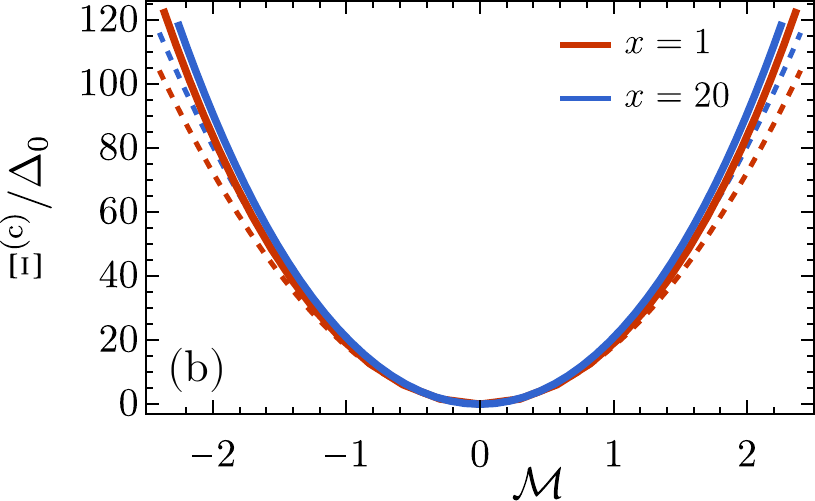}
\caption{The reduced scaling functions $\Xi^{\textrm{(gc)}}/\Delta_0$
(a) and $\Xi^{\textrm{(c)}}/\Delta_0$ (b) for the CCF in the grand canonical and canonical ensembles, respectively. Solid lines indicate the results obtained numerically within nonlinear MFT, while dashed lines show the analytical results of linear MFT as given in Eqs.~\eref{Xigc} and \eref{Xican}. In both cases, the CCF is computed according to Eqs.~\eqref{eq_pCas_pdiff} and \eqref{eq_stressten_dFdL} based on the stress tensor. For illustrative purposes, we have chosen two representative temperatures: $\tscal=1$ (lower, red curves) and $\tscal=20$ (upper, blue curves).  }
\label{fig:scalingfuncs}
\end{figure}

\begin{figure}[t]
\includegraphics[width=.99\columnwidth]{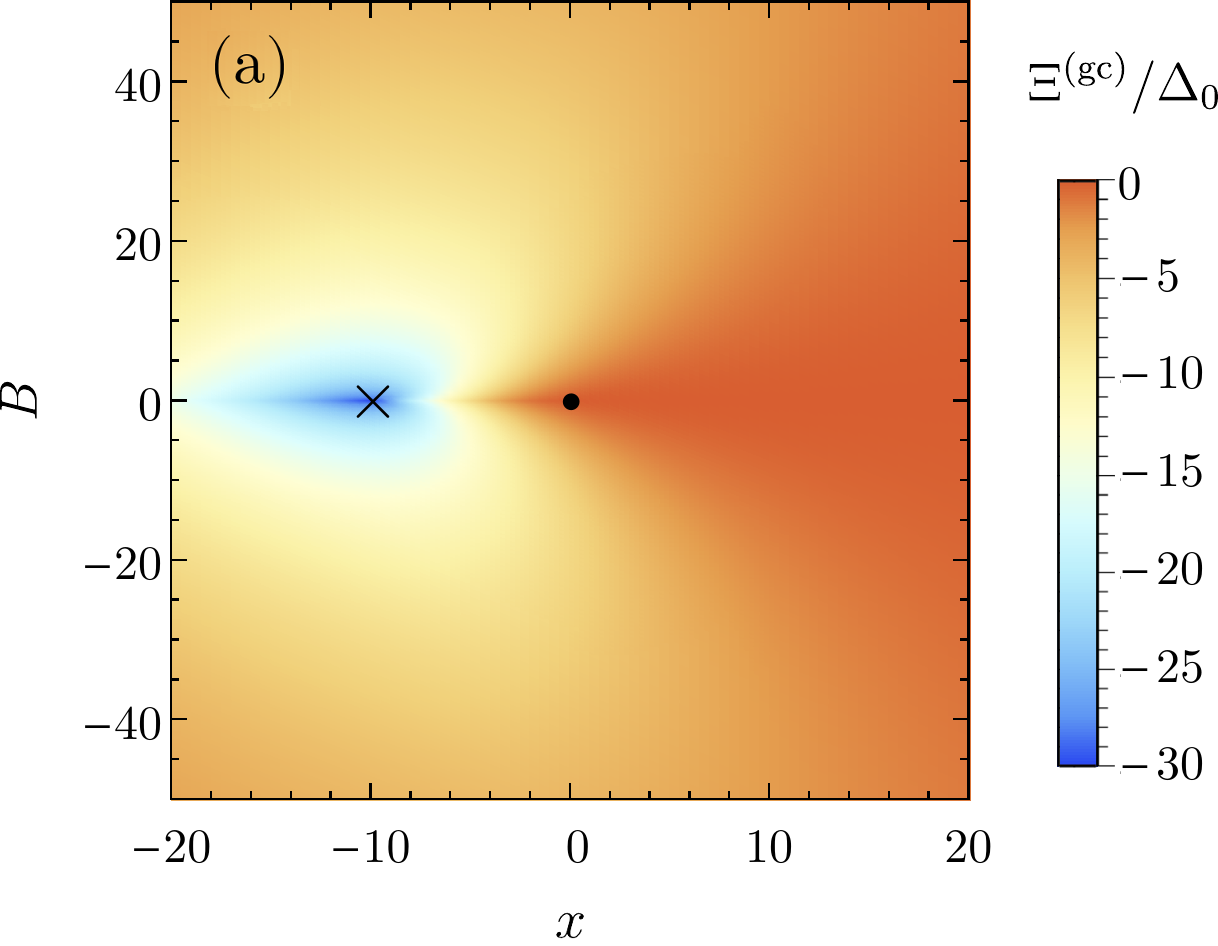} 

\vspace{0.5cm}
\includegraphics[width=.99\columnwidth]{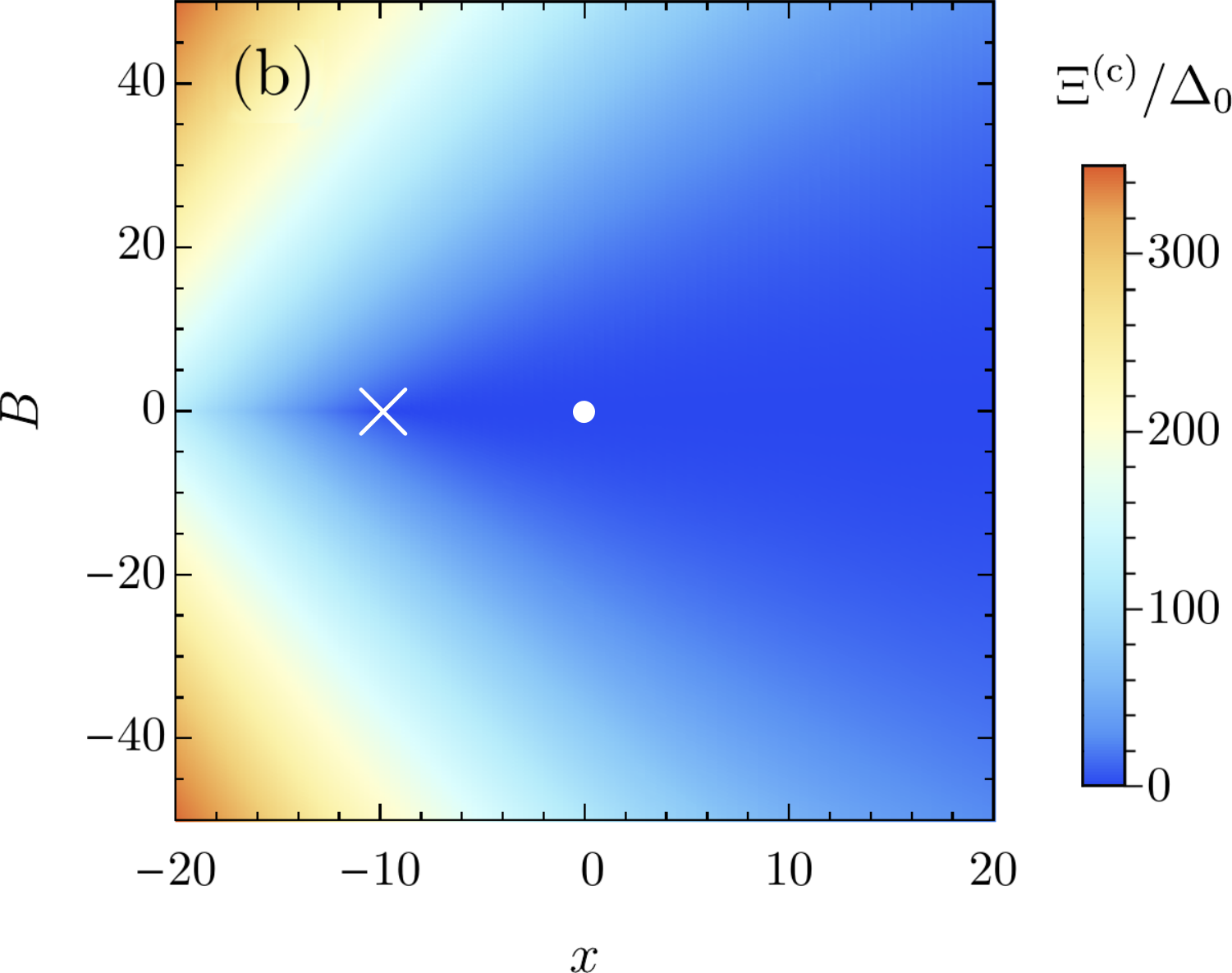}
\caption{The reduced scaling functions $\Xi^{\textrm{(gc)}}/\Delta_0$
(a) and $\Xi^{\textrm{(c)}}/\Delta_0$ (b) for the CCF in the grand canonical and canonical ensemble, respectively, obtained numerically from nonlinear MFT as function of the scaled bulk field $B$ and the scaled temperature $x$. The cross ($\times$) indicates the film critical point ($x=x_c\ut{f}=-\pi^2$, \Eref{eq_x_c_film}), and the dot ($\bullet$) the bulk critical point ($x=0$). In (b), the cross and the dot are in white for better visibility. Note the different color codes. }
\label{fig:CCF_density}
\end{figure}

\begin{figure}[t]
\includegraphics[width=.97\columnwidth]{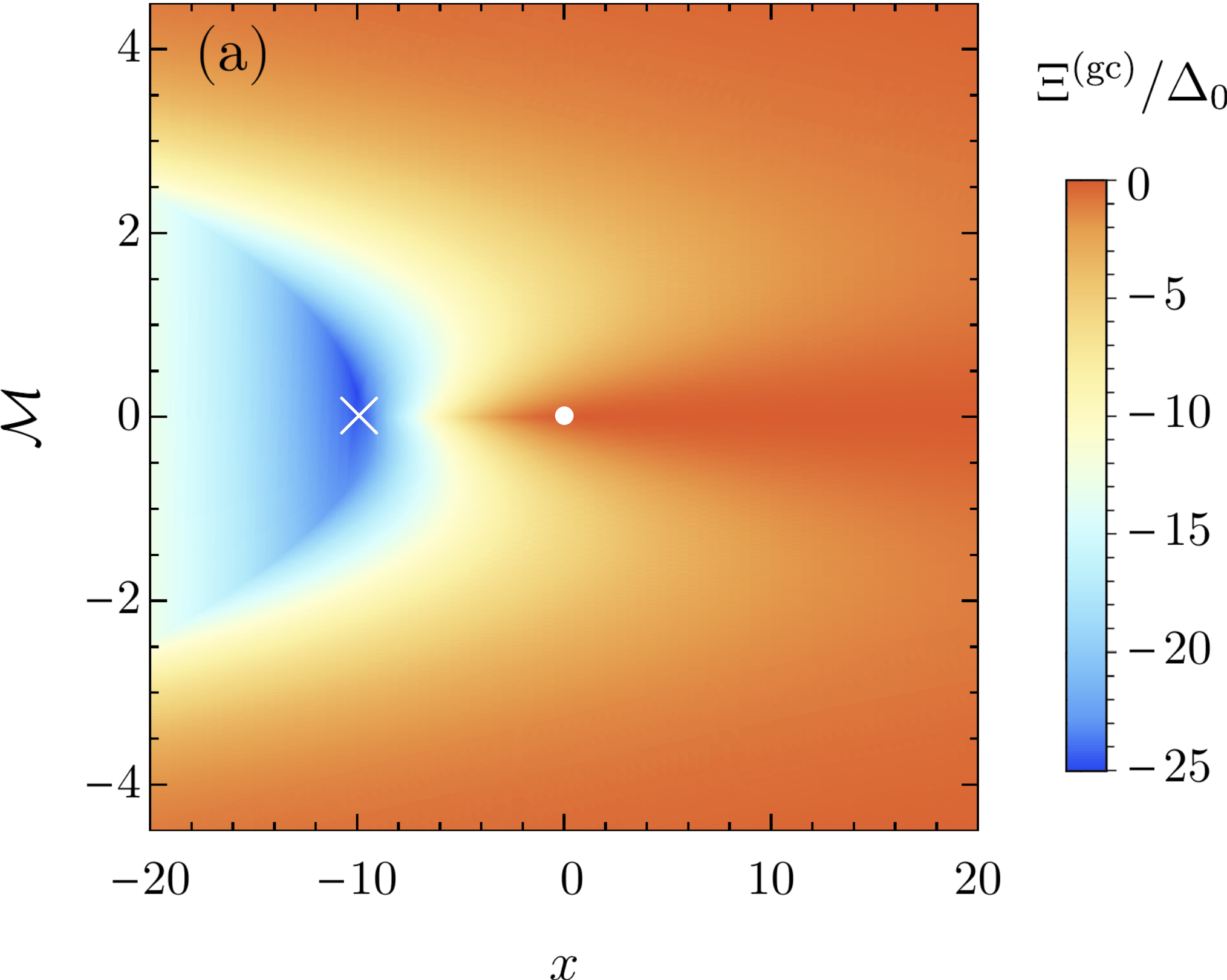} 

\vspace{0.5cm}
\includegraphics[width=.97\columnwidth]{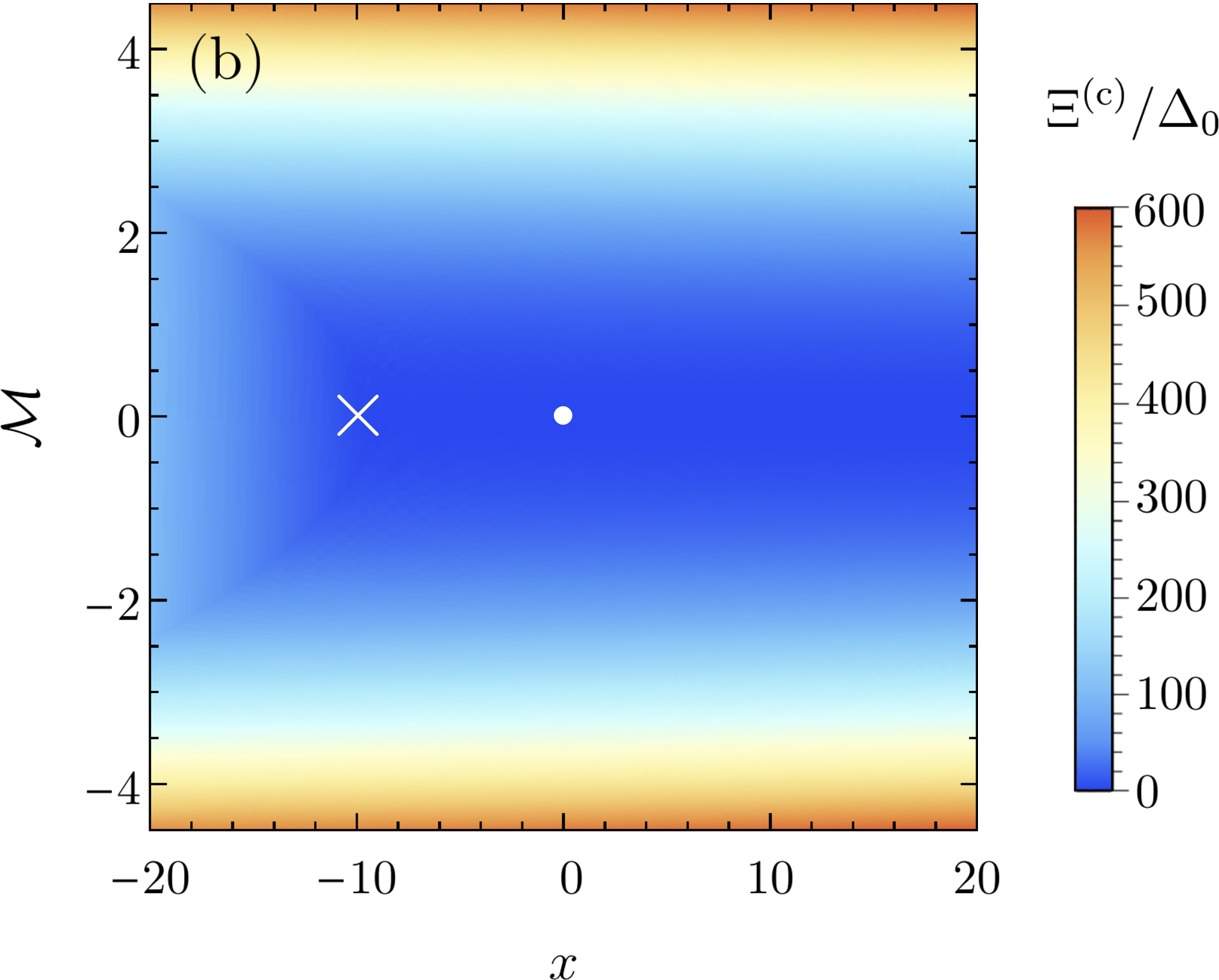}
\caption{The reduced scaling functions $\Xi^{\textrm{(gc)}}/\Delta_0$
(a) and $\Xi^{\textrm{(c)}}/\Delta_0$ (b) for the CCF in the grand canonical and canonical ensemble, respectively, obtained numerically from the nonlinear MFT as function of the scaled mass $\mathcal M$ and scaled temperature $x$. The cross ($\times$) indicates the film critical point ($x=x_c\ut{f}=-\pi^2$, \Eref{eq_x_c_film}), and the dot ($\bullet$) the bulk critical point ($x=0$).}
\label{fig:CCF_density_xM}
\end{figure}

The scaling functions $\Xi\cgc$ of the CCF, as obtained from the stress tensor approach [Eqs.~\eqref{eq_pCas_pdiff} and \eqref{eq_stressten_dFdL}] within nonlinear MFT, are illustrated in \fref{fig:scalingfuncs} (solid lines) as functions of the mass $\Mcal$ for two values of the scaled reduced temperature $x>0$. 
The scaling functions are displayed in reduced form, i.e., divided by the mean-field amplitude $\Delta_0$ [\Eref{eq_Delta0}], which is undetermined within MFT.
Analytical results, obtained within linear MFT and given in Eqs.~\eref{Xigc} and \eref{Xican}, are shown for comparison by the dashed lines. 
In \fref{fig:CCF_density}, the numerically determined scaling functions, obtained within nonlinear MFT, are shown as functions of the scaled bulk field $B$ and of $x$ around the film and the bulk critical point (indicated by the cross and the dot, respectively). 

As illustrated in \fref{fig:scalingfuncs}, linear MFT generally provides an accurate approximation to full MFT for $\Mass\lesssim 1$ and $\tscal\gtrsim 1$.
Notably, within linear MFT and for all values of the reduced temperature $\tscal>0$ and the mass $\Mcal$, the grand canonical CCF reported in \Eref{Xigc} is attractive, i.e., $\Xi\gc<0$, whereas the canonical CCF in \Eref{Xican} is repulsive, i.e., $\Xi\can>0$ \footnote{In order to prove that $\Xi\can>0$, set $y=\sqrt{x}/2$ in \Eref{Xican} and first note that $1-\tanh(y)/y>0$ for $y>0$. Accordingly, $\Xi\can>0$ [\Eref{Xican}] is equivalent to $\tanh(y)> y/(1+y)$. The validity of the latter inequality can be readily shown by considering the derivatives and expressing the hyperbolic functions in terms of exponentials.}.

This character persists also within nonlinear MFT, as can be inferred from \fref{fig:CCF_density}, where the behavior of the CCF scaling functions (determined via the stress tensor approach) as function of the scaled bulk field $B$ and the scaled temperature $x$ is displayed. Figure \ref{fig:CCF_density_xM} presents the same data as function of the scaled mass $\Mcal$ instead of $B$.
In the grand canonical ensemble, generally the CCF is significant only around the film critical point (indicated by a cross in the plots). 
The canonical CCF, in contrast, shows the opposite behavior, growing with increasing distance from the critical region.   

Notably, the difference in sign between the canonical and grand canonical CCF also occurs in the case of critical adsorption with symmetric surface fields (i.e., for $(++)$ boundary conditions) \cite{gross_critical_2016}.
Furthermore, the behavior shown in \fref{fig:scalingfuncs}(a) is consistent with MC results for Ising films in the grand canonical ensemble with a varying bulk field \cite{vasilyev_critical_2013}.

As stated above, the divergence of $\Xi^{\textrm{(gc)}}$ at the critical point is an artifact of linear MFT. Therefore the interval around $\Mcal$, in which the scaling function of linear MFT provides an accurate approximation of the one of nonlinear MFT, becomes progressively narrower upon decreasing $x$. Furthermore, the exact $\Xi^{\textrm{(gc)}}$ is not quadratic in $\mathcal M$, but follows a rather nontrivial form as shown in \fref{fig:scalingfuncs}(a) as well as in Figs.\ \ref{fig:CCF_density} and \ref{fig:CCF_density_xM}. In contrast, as demonstrated in Fig.\ \ref{fig:scalingfuncs}(b), for sufficiently small $|\Mcal|$, $\Xi^{\textrm{(c)}}$ obtained from nonlinear MFT is approximated well by linear MFT, even for $x\to0$.

\subsection{CCF deduced from the free energy}
\label{sec_Forces_FE}

Here we determine the CCF explicitly from the residual finite-size free energy according to \Eref{eq_pCas_dFdL} and compare the result to the one obtained from a pressure difference [\Eref{eq_pCas_pdiff}].

\subsubsection{Grand canonical CCF}

Recalling the lowest order MFT solution in \Eref{eq_phi0z}, we write the grand canonical free energy functional of the film given in \Eref{eq_Landau_func_gc} as
\al{
\mathcal F _f ^{\textrm{(gc)}} &= \int_{0}^{L} dz\;\Big[ \frac 1 2 (\partial_z \phi_0)^2 + \frac 1 2 \tau \phi_0^2 - \mu \phi_0 \Bigr] \nl
&= \underbrace{-\frac{\mu ^2 L}{2 \tau }}_{\textrm{bulk}} + \underbrace{ \frac{\mu ^2}{\tau ^{3/2}}}_{\textrm{surf.}} 
+ \underbrace{ \frac{\mu ^2}{\tau ^{3/2}}  \big[   \tanh \left(\frac{L \sqrt{\tau }}{2}\right) -1   \big] }_{\textrm{residual}} .
}
In the last line we have identified the various contributions according to their scaling with the film thickness $L$, keeping the bare parameters $\tau$ and $\mu$ fixed [see \Eref{eq_Ffilm_split}].
Following \Eref{eq_stressten_dFdL}, the grand canonical film pressure is computed by differentiating the film free energy w.r.t.\ $L$ while keeping the relevant control parameter, in this case $\mu = \tilde \mu$ [see \Eref{eq_mu_constr}], fixed:
\al{
p_f^{\textrm{(gc)}}(\tilde \mu) &=-\partial_L \mathcal F _{f} ^{\textrm{(gc)}}|_{\tilde \mu} \nl
&= \frac{\tau \varphi ^2}{2} \frac{ \tanh ^2\left(\frac{L \sqrt{\tau}}{2}\right)}{ \left(1-\frac{2 \tanh \left(L \sqrt{\tau}/2\right)}{L\sqrt{\tau }}\right)^2}. 
\label{pfgcFreeEn}
}
As expected, \Eref{pfgcFreeEn} is identical to  Eq.~\eref{pfcstresstens}, which was obtained from the stress tensor. On the other hand, the CCF computed via \Eref{eq_pCas_dFdL},
\al{
-\partial_L \mathcal F _{\textrm{res}} ^{\textrm{(gc)}}|_\mu &=-\partial_L \left( \frac{\mu ^2}{\tau ^{3/2}} \tanh \left(\frac{L \sqrt{\tau }}{2}\right)\right) \nl
&= -\frac{\mu ^2}{\tau[1+  \cosh \left(L \sqrt{\tau }\right)]} = \mathcal K^{\textrm{(gc)}}_0 ,
}
is identical to the expression in \Eref{Kgc}. Thus in the grand canonical ensemble, the CCF can be determined equivalently either via the stress tensor or via the residual free energy.

\subsubsection{Canonical CCF}

Inserting the constrained profile (Eq. \eref{eq_phi0ztilde}) into Eq.~(\ref{eq_Landau_func_c}) yields the canonical free energy
\al{
\mathcal F _f ^{\textrm{(c)}} &= \int_{0}^{L} dz\;\Big[ \frac 1 2 (\partial_z \tilde \phi_0)^2 + \frac 1 2 \tau \tilde \phi_0^2 ] \nl
&= \frac{L \tau  \varphi ^2}{2} \Big[\frac{1}{ 1-\frac{2}{L \sqrt{\tau }} \tanh \left(\frac{L \sqrt{\tau }}{2}\right)} \Big] \nl
& = \underbrace{L \tau  \varphi ^2/2}_{\textrm{bulk}} +
\underbrace{\sqrt{\tau } \varphi ^2}_{\textrm{surf.}} +
\underbrace{\sqrt{\tau } \varphi ^2 \big[ \frac{1}{\coth \left(\frac{L \sqrt{\tau }}{2}\right)-\frac{2}{L \sqrt{\tau }}} -1\big]}_{\textrm{residual}} .
\label{FfilmCan}
}
In the last equation, the various contributions have again been identified according to their scaling behavior as function of $L$, keeping the parameters $\tau$ and $\varphi$ fixed, as it is appropriate for a finite-size scaling analysis in the canonical ensemble [see the discussion after \Eref{eq_pB_lim_pF}].
In order to obtain the film pressure via \Eref{eq_stressten_dFdL}, the total mass $\mass$, which is the actual control parameter in the canonical ensemble, is kept fixed, yielding
\al{
p_f^{\textrm{(c)}}&=-\partial_L \mathcal F _{f} ^{\textrm{(c)}}|_\Phi \nl
&= \frac{\tau \varphi ^2}{2} \frac{ \tanh ^2\left(\frac{L \sqrt{\tau }}{2}\right)}{ \left(1-\frac{2 \tanh \left(L \sqrt{\tau}/2\right)}{L\sqrt{\tau }}\right)^2}. 
\label{pfcFreeEn}
}
Since \Eref{pfcFreeEn} and \Eref{pfcstresstens} are equal, herewith the equivalence of computing the film pressure via \Eref{eq_stressten_dFdL} or \Eref{eq_pf_landau} is also established for the canonical ensemble. (Of course equality with the grand canonical film pressure holds, too.)

However, we note that the derivative (at fixed $\mass$) of the residual part of the free energy in \Eref{FfilmCan} is not equal to the canonical CCF in \Eref{Kc} computed via the stress tensor:
\al{
-\partial_L \mathcal F _{\textrm{res}} ^{\textrm{(c)}}|_\Phi &= \Big[ \mathcal K^{\textrm{(c)}}_0 \textrm{ from Eq. \eref{Kc}} \Big] -\frac{2  \varphi ^2 \sqrt{\tau }}{L} \nl
&= \frac{\tau \varphi ^2}{2} \left[\frac{ \tanh ^2\left(\frac{L \sqrt{\tau }}{2}\right)}{ \left(1-\frac{2 \tanh \left(\frac{L \sqrt{\tau }}{2}\right)}{L\sqrt{\tau }}\right)^2}- 1\right]-\frac{2  \varphi ^2 \sqrt{\tau }}{L} \nl
&= \tilde \Kcal_0\can = L^{-d} \tilde\Xi_0\can
\label{eq_CCF_c_Fres}}
with the reduced scaling function
\al{
\frac{\tilde\Xi_0\can}{\Delta_0} &=  \frac{\mathcal M^2 x}{2} \Bigg[\frac{\tanh^2\left(\frac{\sqrt x}{2}\right)}{\left(1-\frac 2 {\sqrt{x}} \tanh \left(\frac{\sqrt{x}}{2}\right)\right)^2}-1 \Bigg] \nl
&\quad - 2  \mathcal M^2 \sqrt{x} .
}
The canonical CCF $\bar\Kcal_0\can$ is still repulsive, but instead of exhibiting the divergence $\propto \sqrt{x}$ in \Eref{Xicxinf}, it attains the finite limit 
\al{ \frac{\tilde\Xi_0\can(x\gg 1)}{\Delta_0} \simeq 6\Mass^2 .
\label{eq_Xic_alt_largeX}}
The term $-\frac{2 \sqrt{\tau } \varphi ^2}{L}$ in \Eref{eq_CCF_c_Fres} would be absent if instead we would compute $-\partial_L \left(\mathcal F _{\textrm{res}} ^{\textrm{(c)}} + \mathcal F _{\textrm{surf}} ^{\textrm{(c)}}\right)\big|_\Phi$, where $\Fcal\can\st{surf}$ denotes the surface contribution identified in \Eref{FfilmCan}. This indicates that the decomposition of $\mathcal F _f ^{\textrm{(c)}} $ according to the standard finite-size scaling arguments in \Eref{FfilmCan} yields a surface contribution which is \emph{not} independent of $L$, and thus contributes to the CCF in the canonical ensemble. This is a genuine consequence of the OP constraint $\mass=\const$. A similar observation has been made for the case of critical adsorption~\cite{gross_critical_2016}.

\section{MC simulations of the Ising model}
\label{sec_MC}

In this section we determine the CCF via MC simulations of the Ising model in a thin film with Dirichlet \bcs in $d=3$ spatial dimensions.
We consider a simple cubic lattice of size $L_{x} \times L_{y} \times L_{z}$ with unit lattice spacing so that $L_x$, $L_y$, and $L_z$ are dimensionless. We apply  periodic boundary conditions along  the  $x$ and $y$ direction and Dirichlet boundary conditions in the $z$ direction. This means, that spins in the bottom layer have no bottom neighbor and spins in the top layer have no top neighbor. At each  lattice site $i=( 1 \le x_i \le L_{x}, 1 \le y_i \le L_{y}, 1 \le z_i \le L_{z})$ a spin $s_i=\pm 1$ is located. 

\subsection{General simulation method}
\label{sec:ising_mag}

In the grand canonical ensemble and in the presence of a uniform bulk field $\mu$, the Hamiltonian of the Ising model for a particular spin configuration $\omega$ is given by 
\begin{equation}
\label{eq:H}
{\cal H}\gc(\omega) = -  J\sum_{\la  ij \ra}  s_{i}  s_{j} -\mu \sum_{ \la k \ra }s_{k}.
\end{equation}
The sum $\la  ij \ra$ is taken  over  nearest neighbors 
 on the lattice and  the sum  $ \la k \ra$ runs  over all spin sites. The energy and the bulk field $\mu$ are
measured in units of the spin-spin interaction constant $J$ so that they become dimensionless and $J=1$.
The grand canonical free energy of the system is
\be
\beta\mathcal F\gc(\beta,\mu)=-\ln \left[\sum \limits_{\{ \omega \}} 
\exp \left( -\beta {\cal H}\gc(\omega) \right) \right],
\ee
where the sum is taken over all spin configurations $\{ \omega \}$;  $\beta=\frac{1}{k_B T}$ denotes 
the inverse thermal energy which in units of $J$ is the dimensionless inverse temperature $\beta=1/T$.
The bulk critical point of the 3d Ising model occurs at the inverse temperature $\beta_{c} \simeq 0.22165455(3)$ \cite{deng_simultaneous_2003}.
We recall that, for a vanishing magnetic field $\mu=0$, the correlation length is [see \Eref{eq_gen_correl}]
$\xi_{t}(t) =\xi_{\pm}^{(0)} t^{-\nu}$ whereas 
at the critical temperature the correlation length is $\xi_{\mu}(\mu) =
\xi_{\mu}^{(0)} |\mu|^{-\nu/\Delta}$
 with the value of the universal correlation length
critical exponent $\nu=0.63002(10)$~\cite{Hasnu},  the universal bulk magnetic field exponent $\Delta=1.5637(14)$~\cite{pelissetto_critical_2002},
and with non-universal critical amplitudes  $\xi_{\mu}^{(0)}=0.278(2)$~\cite{OV}, 
$\xi_{-}^{(0)}=0.243(1)$, and $\xi_{+}^{(0)}=0.501(2)$~\cite{ruge_correlation_1994}. 

\begin{figure}[t]
\includegraphics[width=0.49\textwidth]{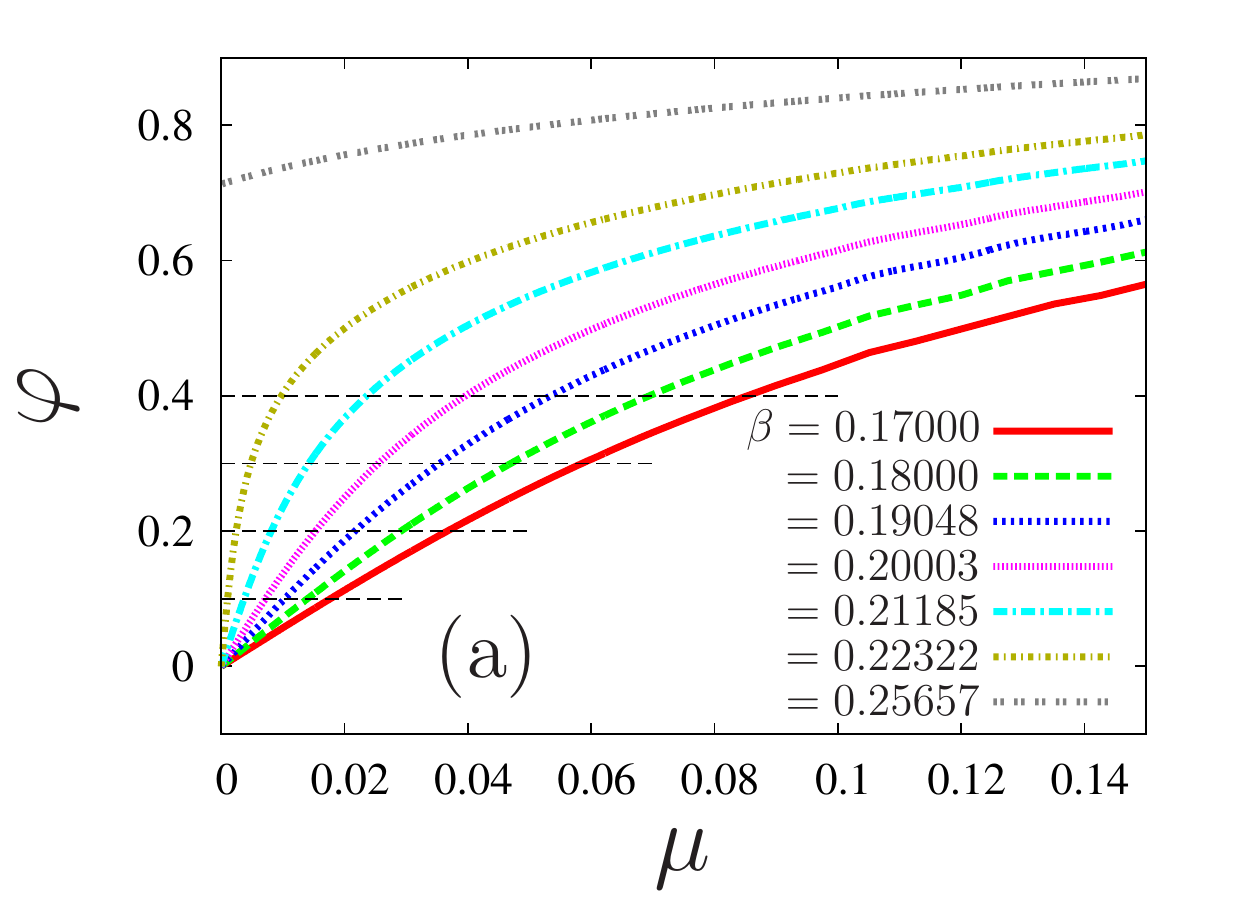}
\includegraphics[width=0.49\textwidth]{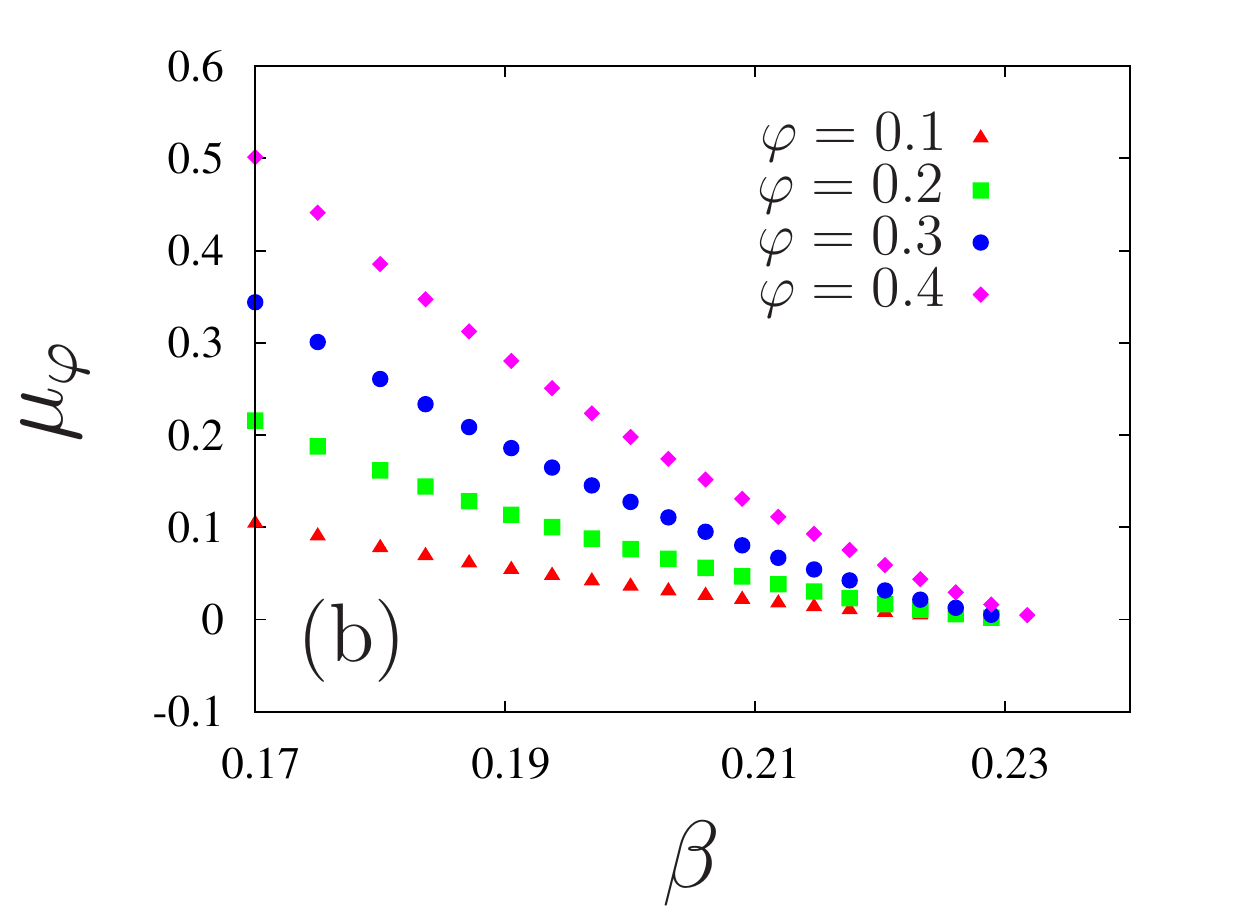}
\caption{MC simulation data (a) The magnetization $\varphi$ per spin as a function of the bulk field $\mu$
for several values of the inverse temperature: $\beta=0.17,0.18,0.19048,0.20003,0.21185,0.22322,0.25657$.
(b) These bulk magnetic fields $\mu_{\varphi}$ as function of the inverse temperature $\beta$ which render the 
values $\varphi=0.1,0.2,0.3,0.4$ of the mean magnetization in the grand canonical ensemble. The four values of $\varphi$ considered in (b) are indicated in (a) by horizontal dotted lines. The value $\mu_\varphi=0$ corresponds to $\varphi=0$.
}
\label{fig:Mbeta}
\end{figure}

The numerical simulation of the Ising model in the grand canonical ensemble has been performed by using a hybrid MC algorithm~\cite{landau_guide_2009}: each MC step consists of a flip of a Wolf cluster followed by $L_{x}\times L_{y} \times L_{z}$ attempts to flip a randomly selected spin 
in accordance with the Metropolis rate.  We perform simulations for a set 
of 32 points $0.17 \le  \beta_{j} \le 0.28 $ with a system of size $60 \times 60 \times 10$, corresponding to an aspect ratio of $L_z/L_{x(y)} \approx 0.167$.
For each value of the inverse temperature $\beta_j$ we have performed 32 simulations, using for each of them a different value $\mu_i$ of the bulk magnetic field with $0 \le \mu_i  \le 0.15$. 
Subsequently, a histogram of the bulk magnetization $\tilde \Phi=\sum \limits_{\la k \ra}  s_{k}$
has been computed for each pair of parameters $(\beta_j, \mu_i)$. The thermal average $\Phi = \bra \tilde \Phi\ket$ has been taken  over $10^{6}$ MC steps, which are split into 10 series in order to assess the numerical error.
We have used the histogram reweighting technique~\cite{landau_guide_2009} in order to compute the mean magnetization
per spin $\varphi=\Phi/(L_{x}L_{y}L_{z})$ [see \Eref{eq_mden}] as a continuous function of the bulk magnetic field $\mu$.
In Fig.~\ref{fig:Mbeta}(a), the magnetization $\varphi$ per spin is shown as function of  $\mu$ 
for several values of the inverse temperature $\beta$.
This information has been used to compute that value $\mu_{\varphi}$ of the bulk magnetic field  which renders the given mean magnetization $\varphi$ per spin for a fixed value of $\beta$. In Fig.~\ref{fig:Mbeta} (b) we plot $\mu_{\varphi}$ as a function of $\beta $ for several values of the magnetization $\varphi$ per spin. These values of $\varphi$ are also indicated in Fig.~\ref{fig:Mbeta}(a) by the horizontal  dotted lines.

\subsection{CCF in the grand canonical ensemble}

\subsubsection{Computation}

The CCF $\mathcal K^{(\textrm{gc})}(\beta,L_{x},L_{y},L)$ in the grand canonical ensemble can be computed on a lattice  with cross-section $L_{x}\times L_{y}$ in terms of the finite difference of the free energies for two distinct slab thicknesses. Here, the actual thickness $L$ considered in the calculation of the CCF is given by $L \equiv L_{z}-\frac{1}{2}$, because it is expressed via the difference of slabs of thickness $L_{z}$ and $L_{z}-1$:
\al{
\label{eq:force}
\mathcal K^{(\textrm{gc})}(\beta,\mu,L)
\equiv &- 
\frac{\beta \Delta\mathcal F^{(\textrm{gc})}(\beta,\mu,L_{x},L_{y},L)}
{L_{x} L_{y}} \nl
&+ \beta f\gc_b(\beta,\mu),
}
where the free energy difference is
\al{
\Delta\mathcal F^{(\textrm{gc})}(\beta,\mu,L_{x},&L_{y},L)= 
\mathcal F^{(\textrm{gc})}(\beta,\mu,L_{x},L_{y},L+\frac{1}{2}) \nl
&-\mathcal F^{(\textrm{gc})}(\beta,\mu,L_{x},L_{y},L-\frac{1}{2}) \, .
}
In the grand canonical ensemble with $\mu \ne 0$, we have computed the free energy difference $\Delta \Fcal$ via the so-called coupling parameter approach. The bulk free energy density
$f_b\gc$ has been computed for the same system but of size $60 \times 60 \times 120$, using the so-called energy integration technique. First, we have computed the bulk free energy at zero bulk field, upon integrating the energy over the inverse temperature. In the next step, for a given value of the inverse temperature, we have integrated the magnetization of the system over the bulk field, obtaining the bulk free energy for a given pair of variables $(\beta_j,\mu_i)$ (see Ref.~\cite{OV} for further details).
In $d$ spatial dimensions the CCF can be expressed in terms of the corresponding scaling function $\Xi\gc$ as
\be
\label{eq:sc}
 \mathcal K^{(\textrm{gc})}(\beta,\mu,L)
=L_{\mathrm{eff}}^{-d}\, \Xi^{(\textrm{gc})}\left(L_{\mathrm{eff}}/ {\xi_{t}},
L_{\mathrm{eff}}/\xi_{\mu}\right).
\ee
We have taken into account finite-size corrections via an effective slab thickness $L_{\mathrm{eff}} \equiv L+ \delta L$, with a correction $\delta L=1.22(2)$ for Dirichlet \bcs \cite{vasilyev_critical_2013}. 

\subsubsection{Discussion}

\begin{figure}[t]
\includegraphics[width=0.97\columnwidth]{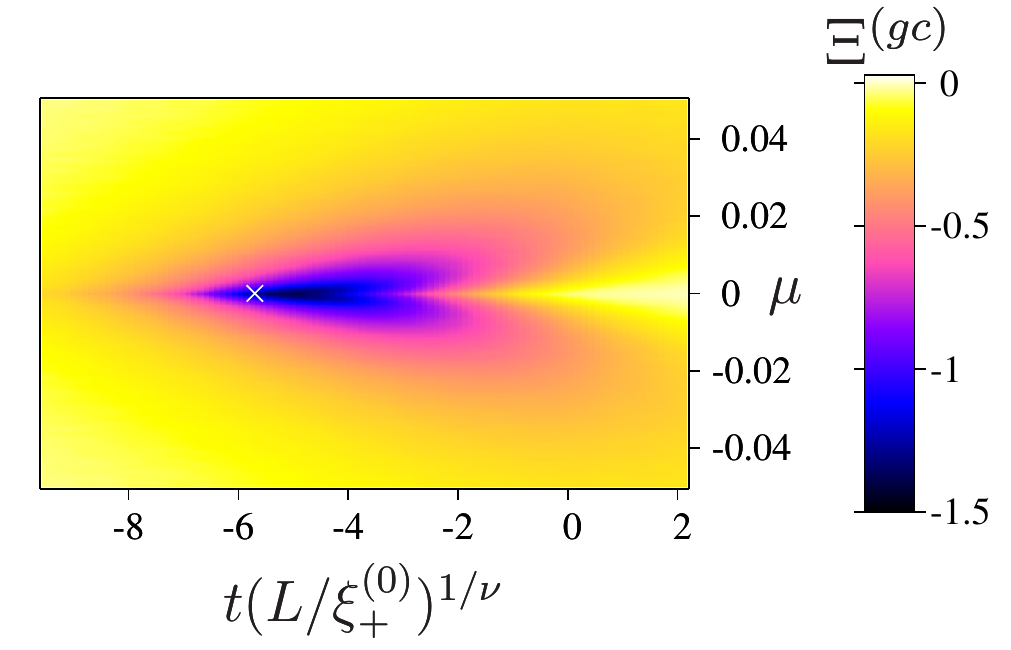}
\caption{
MC results for the CCF scaling function $\Xi^{(\textrm{gc})}$ for the grand canonical  ensemble as function of the temperature scaling variable $t(L/\xi_+^{(0)})^{1/\nu}$ and the bulk magnetic field $\mu$. The cross ($\times$) indicates the film critical point of the Ising model \cite{vasilyev_universal_2009}.
}
\label{fig:fc3d}
\end{figure}
Fig.~\ref{fig:fc3d} shows the scaling function of the grand canonical CCF obtained from our MC simulations as function of the temperature scaling variable $t(L/\xi_+^{(0)})^{1/\nu}=x$ [\Eref{eq_tscal}] and of the bulk magnetic field $\mu$. 
The behavior of $\Xi\gc$ obtained within MFT [see  Fig.~\ref{fig:CCF_density}(a)] qualitatively agrees with our MC simulations. 
Consistently with previous studies \cite{vasilyev_critical_2013}, we find that the grand canonical CCF is attractive and reaches its greatest strength at vanishing bulk field $\mu=0$ and at a slightly negative reduced temperature $t<0$. 
However, a more quantitative comparison is precluded due to appearance of the undetermined amplitude $\Delta_0$ [\Eref{eq_Delta0}] arising within MFT.
This deficiency can be overcome by including fluctuation effects within a renormalization group approach \cite{dohm_pronounced_2014}.

\begin{figure}[t]
\includegraphics[width=0.49\textwidth]{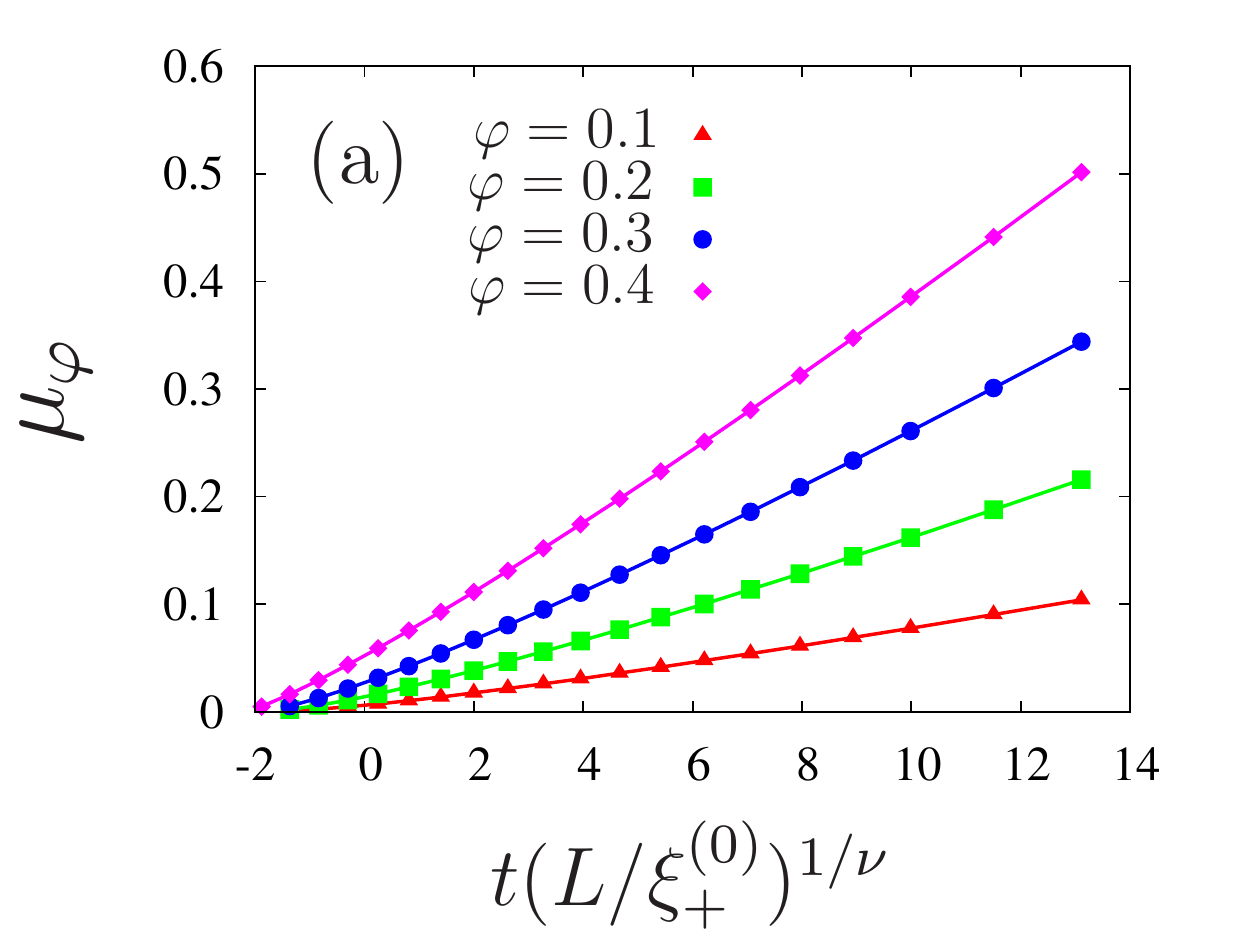}
\includegraphics[width=0.49\textwidth]{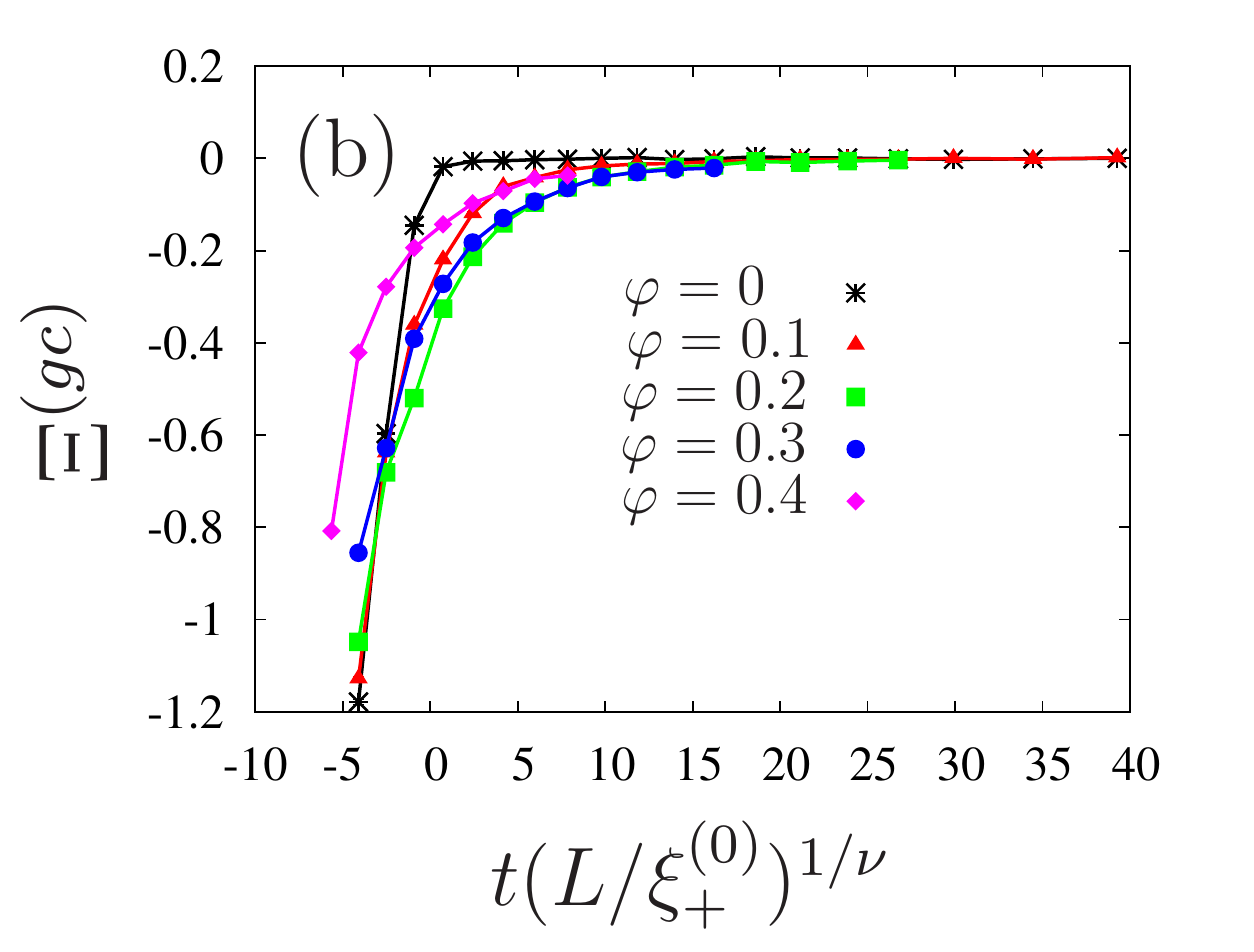}
\caption{MC results in the grand canonical ensemble.
(a) Value $\mu_{\varphi}$ of the bulk magnetic field,  at which the mean magnetization takes the values $\varphi=0.1,0.2,0.3,0.4$, as a function of the temperature scaling variable $t(L/\xi_+^{(0)})^{1/\nu}$. In the considered temperature region, $\varphi=0$ corresponds to $\mu_\varphi=0$.
(b) The scaling function $\Xi^{(\textrm{gc})}$ of the grand canonical CCF as a function of the temperature scaling variable $t(L/\xi_+^{(0)})^{1/\nu}$, computed along the lines $(t(L/\xi_+^{(0)})^{1/\nu},\mu_{\varphi})$ in (a) for the mean magnetizations $\varphi=0, 0.1,0.2,0.3,0.4$. 
}
\label{fig:hfc}
\end{figure}

In Fig.~\ref{fig:hfc}(a) we illustrate the relationship between $\mu$ and the scaling variable $t(L/\xi_0^+)^{1/\nu}$ for various values of the mean magnetization $\varphi$.
\Fref{fig:hfc}(b) shows the CCF scaling function
$\Xi^{(\textrm{gc})}$ along lines of fixed magnetization $\varphi$ as a function of $t(L/\xi_0^+)$. 
The CCF for $\varphi=0$ --- the data of which have been presented previously in Ref.\ \cite{vasilyev_universal_2009} --- is weak and attractive and an accurate, corresponding field theoretic description has been provided in Ref.\ \cite{dohm_pronounced_2014}.
Th representation of the CCF in \Fref{fig:hfc}(b) allows one to directly compare the grand canonical results with those in the canonical ensemble, to which we turn next.

\subsection{CCF in the canonical ensemble}
\subsubsection{Computation}

The Hamiltonian of the Ising model for the canonical ensemble is given by
\begin{equation}
\label{eq:HC}
{\cal H}\can(\omega) = -  J\sum_{\la  ij \ra}  s_{i}  s_{j}
\end{equation}
and does not include the bulk magnetic field. The canonical free energy is obtained from the partition function as
\al{
\beta\mathcal F\can(\beta,m)=-\ln \bigg[&\sum \limits_{\{ \omega \}}
\delta( N m, \sum \limits_{\la k \ra} s_k(\omega) )  \nl
&\times \exp \left( -\beta {\cal H}\can(\omega) \right) \bigg],
}
where the Kronecker delta function selects only terms corresponding to spin configurations $\omega$  with fixed, prescribed magnetization $\Phi= \varphi N= \sum \limits_{\la k \ra} s_k(\omega) $.
Here, $N=L_x \times L_y \times L_z$ denotes the total number of spins in the system.
A configuration of $N$ spins with a magnetization $\varphi$ per spin contains  $N_+=\frac{1+\varphi}{2}N$ up spins and $N_-=\frac{1-\varphi}{2}N$ down spins. 
At infinite temperature $\beta=0$, the free energy of the canonical ensemble can be expressed as
\be
\beta \mathcal F\can( \beta,\varphi,N)\big|_{\beta=0}=-\ln \left( \frac{N!}{N_{+}! N_{-}!} \right).
\ee
Using Stirling's formula $n! \simeq\sqrt{2 \pi } n^{n+\frac{1}{2}}\exp(-n)$, one obtains for the bulk free energy per spin $\beta f\can_b=\frac{1}{N} \beta \mathcal F_b^{(\textrm{c})}( \beta,m,N)$ in the limit of high temperatures ($\beta=0$) and large system sizes $N$
\al{
\label{eq:fbC}
\beta f\can_b(\beta,\varphi)|_{\beta=0}&=\frac{1+\varphi}{2}\ln\left( \frac{1+\varphi}{2}\right)  +\frac{1-\varphi}{2}\ln\left( \frac{1-\varphi}{2}\right) \nl
&\quad + \frac{1}{2N} \ln\left(\frac{\pi}{2}(1+\varphi)(1-\varphi) \right).
}
We note that in principle the canonical free energy density in Eq.~(\ref{eq:fbC})  differs from the grand canonical free energy density, which at infinite temperature is $\beta f_b\gc(\beta)|_{\beta=0}=- \ln(2)$.
Only for zero magnetization $\varphi=0$ these two quantities coincide, {\it i.e.}, $\beta f\can_b(\beta,\varphi=0)|_{\beta=0}=- \ln(2)$.

For the canonical ensemble we have computed the free energy for a system with cross-section $L_{x}\times L_{y}$ and thickness $L_{z}$ ($N=L_x \times L_y \times L_z$) via integration of the mean energy $E$ per spin over the inverse temperature $\beta$:
\begin{equation}
\label{eq:FC}
\beta\mathcal F^{(\textrm{c})}(\beta,\varphi,L_{z})
=-\ln \left( \frac{N!}{N_{+}! N_{-}!} \right)+
\int \limits_{0}^{\beta}  E\can(\beta',\varphi,L_z) 
 {\mathrm d }\beta'.
\end{equation}
The internal energy $E\can$ of the canonical system with a fixed magnetization $\varphi$ per spin has been computed based on Kawasaki dynamics~\cite{kawasaki_diffusion_1966}. Typically we have used $5 \times 10^{7}$ MC steps for thermalization (one MC step consists of $N$ attempts of pair Kawasaki exchanges), followed by $10^8$ MC steps for computing the thermal average. Using Eq.~(\ref{eq:FC}) we have determined the free energy difference
$\Delta\mathcal F^{(\textrm{c})}(\beta,\varphi,L_{x},L_{y},L)=
\mathcal F^{(\textrm{c})}(\beta,\varphi,L_{x},L_{y},L+\frac{1}{2})-
\mathcal F^{(\textrm{c})}(\beta,\varphi,L_{x},L_{y},L-\frac{1}{2})$.
Without knowledge of the bulk free energy density $f\can_b$ we can apply the method introduced in Ref.~\cite{vasilyev_monte_2007}, which provides the following difference:
\al{
\label{eq:gc}
g_{C}(\beta,\varphi,L)&=\beta \big[ \Delta\mathcal F^{(\textrm{c})}(\beta,\varphi,L_{x},L_{y},L) \nl
&\quad -\Delta \Fcal\can(\beta,\varphi,L_{x},L_{y},2L) \big].
}
Considering therein the second term as an estimate of a difference of the bulk free energy, we approximate the canonical CCF as $\Kcal\can(\beta,\varphi,L)\approx g_C(\beta,\varphi,L)$.
Accordingly, the associated scaling function follows, analogously to \Eref{eq:sc}, as
\be
\label{eq:thetaC}
\Xi^{(\textrm{c})} \left( L_{\mathrm{eff}}/{\xi_{t}},\varphi\right) 
\simeq L_{\mathrm{eff}}^d g_{C}(\beta,\varphi,L).
\ee
In order to numerically determine $\Xi\can$ via Eq.~(\ref{eq:thetaC}), we have performed MC simulations for a system of size $L_x=L_y=60$, $L_z=10$, and we have assumed an effective thickness 
$L_{\mathrm{eff}}=L+1.22$ as in the grand canonical case. 

\subsubsection{Discussion}

\begin{figure}[t]
\includegraphics[width=0.49\textwidth]{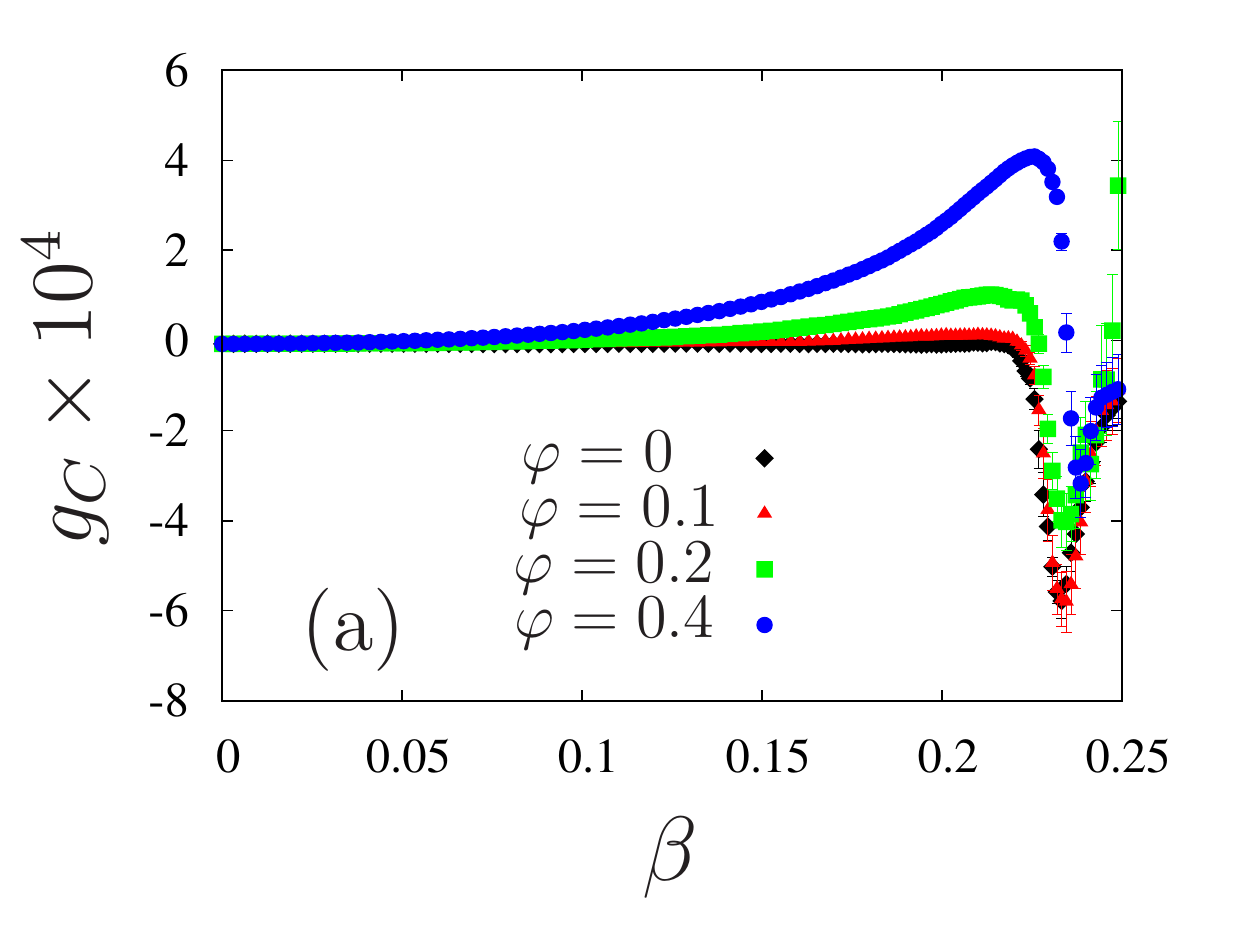}
\includegraphics[width=0.49\textwidth]{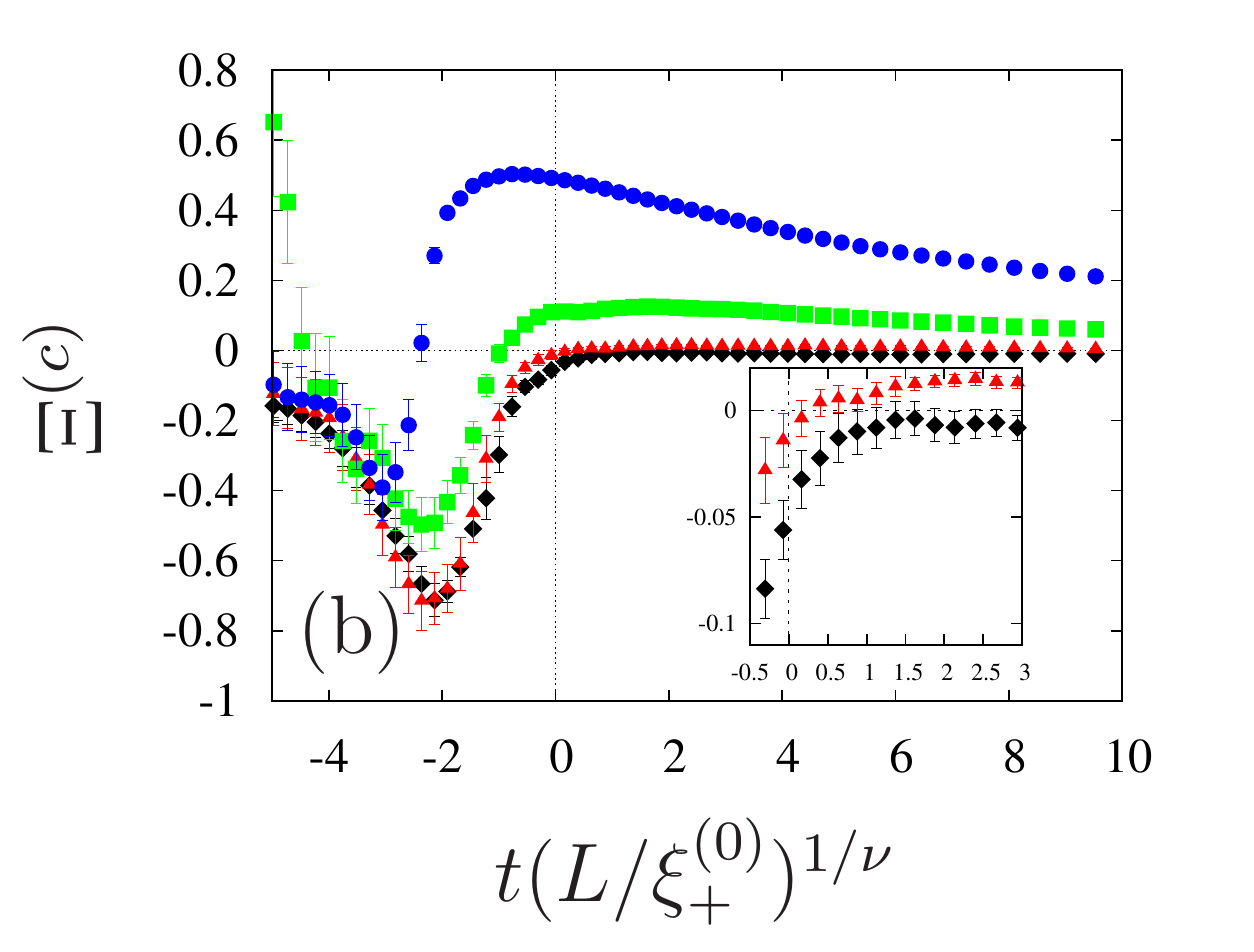}
\caption{MC results in the canonical ensemble. (a) Auxiliary function $g_C$ [Eq.~(\ref{eq:gc})] as a function of the inverse temperature $\beta$ for $\varphi=0.1,0.2,0.4$.
(b) Scaling function $\Xi^{(\textrm{c})}$ of the canonical CCF [\Eref{eq:thetaC}] as function of the scaled temperature $t(L/\xi_+^{(0)})^{1/\nu}$ for three mean magnetizations $\varphi=0,0.1,0.2,0.4$.}
\label{fig:gc}
\end{figure}

In Fig.~\ref{fig:gc}(a) the auxiliary function $g_C$ [Eq.(\ref{eq:gc})] is plotted as a function of $\beta$ for $\varphi=0,0.1,0.2,0.4$. 
\Fref{fig:gc}(b) shows the scaling function $\Xi^{(\textrm{c})}$ of the canonical CCF obtained via \Eref{eq:thetaC} as a function of the scaled temperature $t(L/\xi_+^{(0)})^{1/\nu}$ for various mean magnetizations $\varphi>0$.
In contrast to the grand canonical ensemble [see Fig.~\ref{fig:hfc}(b)], the canonical CCF obtained from  MC simulations is repulsive ({\it i.e.}, $\Xi\can>0$) for $\varphi\gtrsim 0.1$ and for supercritical temperatures $x>0$.
The repulsive character and the fact, that the strength of the canonical CCF increases with growing $|\varphi|$, are captured correctly by MFT [see Fig.~\ref{fig:CCF_density_xM}(b)].   
Note, however, that, within MFT, the canonical CCF is repulsive across the whole parameter range considered here, except at $\varphi=0$, where $\Xi\can=0$.
The discrepancy between MC results and MFT concerning the sign of the CCF for $t>0$ and as $\varphi\to 0$ is due to the effect of critical fluctuations, which render a weak but attractive canonical CCF, in agreement with the predictions in Ref.\ \cite{gross_statistical_2017}, \footnote{The high-temperature limit of $\Xi\can$ for $\varphi=0$ obtained from MC simulations differs from the theoretical expectation $\Xi\can\simeq -\varrho^{d-1}/2$ (where $\varrho=L_z/L_{x,y}$ is the aspect ratio) mainly due to the approximations involved in \Eref{eq:gc}}.
A detailed analysis of the CCF for subcritical bulk temperatures ($t<0$), for which, in contrast to the predictions of MFT, a pronounced minimum appears, is left for future studies.

\section{Summary}

We have studied ensemble differences of the OP profile and of the CCF, arising in a critical fluid film of thickness $L$ within the so-called ordinary surface universality class at both walls.
In the grand canonical ensemble, the film can exchange material with its environment at a common chemical potential $\mu$.
In the canonical ensemble, instead, particle exchange is prohibited and the film and the environment are taken to have the same mean OP density $\mden$.
The system is analyzed within mean field theory, i.e., neglecting thermal fluctuations. In this limit, the CCF stems solely from the action of an external bulk field (such as the chemical potential $\mu$) or, correspondingly, from a nonzero total mass $\mass=\int_0^L d z\, \phi(z) = L \mden$. We generally assume translational invariance in the lateral directions of the film and, accordingly, we consider all extensive quantities as defined \emph{per transverse area $A$} [compare \Eref{eq_Mass0}].

Our findings can be summarized as follows:
\begin{enumerate}
 \item We have solved the Euler-Lagrange equations for the OP profile in the film via three complementary approaches: (i) a perturbative solution in terms of orders of the nonlinear term in the supercritical regime ($T\geq T_c\ut{b}$), (ii) an exact solution below the film critical point, i.e., for $T\leq T_c\ut{f}<T_c\ut{b}$ and at vanishing external bulk field ($\mu=0$), and (iii) a numerical solution for arbitrary values of the temperature and of the bulk field. For small values of the scaled mass $|\Mass|$ or, respectively, the scaled bulk field $|B|$, the perturbative solution generally provides an accurate approximation of the exact mean field solution [see Figs.~\ref{exact-num}, \ref{uncontrained_profiles}, and \ref{constrained_profiles}].
 
 \item The scaling behavior of the scaled mass $\Mass(\tscal,B)$ as function of the scaled temperature $\tscal$ and the scaled bulk field $B$ has been analyzed based on the full MFT in various asymptotic limits  [see Figs.~\ref{phasediagsuperimp}, \ref{fig:M(x=0,B)}, and \ref{fig:M(x,B=0)}). In the case of a vanishing bulk field ($B=0$), the exact expression for $\Mass(\tscal)$ has been determined in \Eref{mass_02}. For general $B$, we have explicitly demonstrated that Widom's scaling hypothesis applies for the system studied here [see Fig. \ref{WidomFig}].

 \item We have analyzed the CCF within linear and nonlinear MFT in the canonical and the grand canonical ensembles. For $\Mcal=0$ and $x>0$, both the canonical and the grand canonical CCF vanish within MFT. 
 For $\Mcal\neq 0$ and within the studied parameter region around the film and bulk critical points [see Figs.\ \ref{fig:scalingfuncs}, \ref{fig:CCF_density}, and \ref{fig:CCF_density_xM}], the grand canonical CCF is \emph{attractive} (consistent with Ref.\ \cite{vasilyev_critical_2013}), while the canonical CCF is \emph{repulsive}.

 \item The canonical CCF depends on whether it is defined as the difference between the film and the bulk pressure [\Eref{eq_pCas_pdiff}] or as a derivative of the residual finite-size free energy [\Eref{eq_pCas_dFdL}]. The difference between the two approaches stems from a surface-like pressure contribution [see \Eref{eq_CCF_c_Fres}], which is a direct consequence of the global OP constraint in the canonical ensemble. 
 
 \item The grand canonical CCF generally vanishes in the limit $\tscal\to\infty$, as expected. In contrast, the canonical CCF defined as a pressure difference [\Eref{eq_pCas_pdiff}] diverges $\propto \tscal$ in this limit [see \Eref{Xicxinf}], while the canonical CCF extracted from the residual finite-size free energy [\Eref{eq_pCas_dFdL}] approaches a nonzero constant [see \Eref{eq_Xic_alt_largeX}]. These unexpected limiting behaviors are again induced by the OP constraint acting in the canonical ensemble.
 
 \item We have studied the influence of fluctuations onto the CCF via MC simulations of the $d=3$ dimensional Ising model. The predictions of MFT are qualitatively recovered by the simulations for sufficiently large values of the magnetization $\varphi$, while fluctuations dominate for $|\varphi| \ll 1$. In agreement with MFT, we find that the grand canonical CCF is attractive within the studied parameter ranges of $x$ and $\varphi$ [see Figs.\ \ref{fig:fc3d} and \ref{fig:hfc}(b)]. At supercritical temperatures ($t>0$), the canonical CCF is repulsive for large mean magnetizations $|\varphi| \gg 1$ and (weakly) attractive for small mean magnetizations [see Fig. \ref{fig:gc}(b)]. Below the bulk critical point, a pronounced minimum of the canonical CCF is observed. Note that a quantitative comparison between MFT and MC simulations is precluded by the undetermined amplitude $\Delta_0$ [\Eref{eq_Delta0}] appearing in the mean-field scaling functions.  
 
\end{enumerate}

We remark that certain characteristic features of the canonical CCF found here, such as its repulsive character, its dependence on the precise definition [i.e., \Eref{eq_pCas_dFdL} vs.\ \Eref{eq_pCas_pdiff}], and its nontrivial decay behavior for thick films ($\tscal\gg 1$), appear analogously also in the case of critical films confined by walls with parallel surface fields \cite{gross_critical_2016}. 

Critical fluids typically show strong adsorption at the container walls \cite{liu_universal_1989, floter_universal_1995, gambassi_critical_2009}.
In order to experimentally study the results obtained here, it would thus be necessary to suitably modify the walls in order to obtain effective Dirichlet \bcs for the OP. As has been shown previously, this can be achieved by endowing the surfaces with narrow chemical stripes of antagonistic character \cite{nellen_tunability_2009, sprenger_forces_2006, trondle_normal_2009, trondle_critical_2010, gambassi_critical_2011}.
Together with Refs.\ \cite{gross_critical_2016, gross_statistical_2017}, the present study provides further evidence that the CCF crucially depends on the thermodynamic ensemble under consideration and, in particular, on the presence of OP constraints.

\appendix

\begin{widetext}
\section{Solution of \Eref{ELE_MFT} at $\mathcal O (\epsilon^2)$}
\label{app1}
In terms of the formal expansion of the OP $m$ and of the bulk field $B$ (see \Eref{epsexpand}), at $\Ocal(\epsilon^2)$ the ELE [\Eref{ELE_MFT}] reads
\al{
m''_2 = x m_2 + 3 m_0^2 m_1  - B_2.
\label{app_ELE2}}
While the full solution of \Eref{app_ELE2} is omitted here, we report the expression for the constraint-induced field $B_{2} = \tilde B_2$:
\al{
\tilde B_2 = &\frac{\mathcal M^5 x^{3/2} }{15360 \left(\sqrt{x} \cosh \left(\frac{\sqrt{x}}{2}\right)-2 \sinh \left(\frac{\sqrt{x}}{2}\right)\right)^7} \nl
&\times \Bigg[-2025 \sqrt{x} (8 x+121) \sinh \left(\frac{\sqrt{x}}{2}\right)-216 \sqrt{x} (75 x+1256) \sinh \left(\frac{3 \sqrt{x}}{2}\right) \nl
&-24 \sqrt{x} (135 x+2518) \sinh \left(\frac{5 \sqrt{x}}{2}\right)-681 \sqrt{x} \sinh \left(\frac{7 \sqrt{x}}{2}\right)+50 (4410 x-3659) \cosh \left(\frac{\sqrt{x}}{2}\right)\nl
&+648 (160 x+171) \cosh \left(\frac{3 \sqrt{x}}{2}\right)+40 (432 x+1685) \cosh \left(\frac{5 \sqrt{x}}{2}\right)+4742 \cosh \left(\frac{7 \sqrt{x}}{2}\right)\Bigg],
}
which has the following asymptotic scaling behavior:
\al{
\tilde B_2 \to
\begin{cases}
   -\frac{8343 }{175175} \mathcal M^5,\quad&x\to 0,\\
 \sim -\frac{227}{80}\mathcal{M}^{5}x^{-3/2}\to 0,\quad&x\to\infty.
\end{cases}
}
The constrained profile scales as
\al{
\tilde m_2(\z) \to
\begin{cases}
 p^{(14)}(\zeta-1/2),\quad&x\to 0,\\
 0,\quad&x\to\infty,
\end{cases}
}
where $p^{(14)}$ represents the 14th-order polynomial
\al{
p^{(14)}(y) = -\frac{81 \mathcal M^5}{1435033600}  \Big[&40550400 y^{14}-106229760 y ^{12} +134278144 y ^{10}\nl
&-91703040 y ^8+35939904 y ^6-8408400 y ^4  +959492 y ^2-25365\Big].
}

\end{widetext}

\bibliography{bibliography}

\end{document}